%% file: plurality.TEAC_new.tex
\documentclass[letterpaper]{article}
\usepackage{times}
\usepackage{helvet}
\usepackage{courier}
\usepackage{mathrsfs}
\usepackage{graphicx}
\usepackage{amsmath}
\usepackage{amssymb}
\usepackage{amsthm}
\usepackage{algorithm}
\usepackage{algpseudocode}
\usepackage{multirow} 
\usepackage[square]{natbib}
\usepackage[table]{xcolor}
\usepackage{url}
\usepackage{enumitem}
\usepackage{tkz-graph}
\usepackage[position=top]{subfig}
\let\chapter\section
\input{defs.tex}

\newtheorem{nlemma}[theorem]{Lemma}
\newtheorem{nproposition}[theorem]{Proposition}
\renewcommand{\vec}[1]{\mathbf{#1}}
\newcommand{\hatv}[1]{\hat{\vec{#1}}}

\renewcommand{\ord}[1]{\succ_{#1}}
\newcommand{\jr}[1]{#1}
\newcommand{\ssection}[1]{\vspace{-1mm}\section{#1}\vspace{-1mm}}

\begin{document}
\title{Acyclic Games and Iterative Voting
\jr{\thanks{A preliminary version of this paper has been presented at AAAI-2010~\cite{MPRJ:2010:AAAI}.}
}
}

\author{Reshef Meir$^1$
 \and Maria Polukarov$^2$ \and  Jeffrey S. Rosenschein$^3$ \and Nicholas R. Jennings$^4$  \\ 
$^1$Technion---Israel Institute of Technology \texttt{reshefm@ie.technion.ac.il} \\
$^2$University of Southampton, United Kingdom \texttt{ mp3@ecs.soton.ac.uk}\\
$^3$The Hebrew University of Jerusalem, Israel \texttt{jeff@cs.huji.ac.il} \\
$^4$Imperial College, United Kingdom \texttt{n.jennings@imperial.ac.uk}}



\newcount\Comments  
\Comments=0
\definecolor{darkgreen}{rgb}{0,0.6,0}
\newcommand{\kibitz}[2]{\ifnum\Comments=1{\color{#1}{#2}}\fi}
\newcommand{\rmr}[1]{\kibitz{blue}{[RESHEF:#1]}}
\newcommand{\mpl}[1]{\kibitz{red}{[MARIA:#1]}}

\pagestyle{plain}

\newcommand{\smallcaption}[1]{\caption{\small #1}}

\date{}
\maketitle


\begin{abstract}
\rmr{abstract will depend on the journal}

We consider iterative voting models and position them within the general framework of acyclic games and game forms. More specifically, we classify convergence results based on the underlying assumptions on the agent scheduler (the order of players) and the action scheduler (which better-reply is played).

Our main technical result is providing a complete picture of conditions for acyclicity in several variations of Plurality voting. In particular, we show that (a) under the traditional lexicographic tie-breaking, the game converges for any order of players under a weak restriction on voters' actions; and (b)  Plurality with randomized tie-breaking is not guaranteed to converge under arbitrary agent schedulers, but from any initial state there is \emph{some} path  of better-replies to a Nash equilibrium. We thus show a first separation between restricted-acyclicity and weak-acyclicity of game forms, thereby settling an open question from \cite{kukushkin2011acyclicity}. In addition, we refute another conjecture regarding strongly-acyclic voting rules.
\end{abstract}


\ssection{Introduction}\label{sec:intro}

Researchers in economics and game theory since Cournot~\shortcite{cournot1838recherches} had  been developing a formal framework to study questions about acyclicity and convergence of local improvement dynamics in games.


Intuitively put, strong-acyclicity means that the game will converge regardless of the order of players/voters and how they select their action (as long as the moving agents are improving their utility in every step), i.e. that there are no cycles of better-replies whatsoever; Weak-acyclicity means that while cycles may occur, from any initial state (voting profile) there is at least one path of better-replies that leads to a Nash equilibrium; Restricted-acyclicity is a middle ground, requiring convergence for any order of players (agent scheduler), but allowing the action scheduler to restrict the way they choose among several available replies (e.g., only allowing best-replies). Most relevant to us is the work of Kukushkin~\shortcite{kukushkin99,kukushkin02,kukushkin2011acyclicity}, who studied general characterizations of game forms that guarantee various notions of acyclicity. 

A more recent field is \emph{iterative voting}.
In the iterative voting model, voters have fixed preferences and start from some announcement (e.g., sincerely report their preferences). Votes are aggregated via some predefined rule (e.g. Plurality), but can change their votes after observing the current announcements and outcome. 
The game proceeds in turns, where a single voter changes his vote at each turn, until no voter has objections and the final outcome is announced.  This process is similar to online polls via Doodle or Facebook, where users can log-in at any time and change their vote. Similarly, in offline committees the participants can sometimes ask to change their vote, seeing the current outcome. 

The formal study of iterative voting rules was initiated about 6 years ago in a AAAI paper that was a preliminary version of this one~\cite{MPRJ:2010:AAAI}. Iterative voting papers typically focus on common voting rules such as Plurality and Borda, and study the conditions under which convergence of the iterative process to a Nash equilibrium is guaranteed. Most results in the field consider best-reply dynamics~\cite{lev2012convergence,reyhani2012best,obraztsova2015convergence}. 

While voting rules and game forms are essentially the same thing, the iterative voting literature has remained largely detached from the more general literature on acyclicity in games. Bridging this gap is the main conceptual contribution of this work, for two reasons. First, understanding the conditions that entail acyclicity of games and game forms is crucial to the understanding of iterative voting scenarios, and to properly compare convergence results (e.g. convergence of best-reply dynamics is a special case of restricted acyclicity). Likewise, convergence results for  specific voting rules under best/better-reply dynamics may shed light on more general questions regarding acyclicity.
Building on the formalism of Kukushkin~\shortcite{kukushkin2011acyclicity} for strong/ restricted/ weak-acyclicity of game forms, we re-interpret in this paper both known and new results on convergence of better- and best-reply in voting games, and answer some open questions.



\subsection{Related work}
\label{sec:related}
Kukushkin~\shortcite{kukushkin2011acyclicity} provided several partial characterizations for game forms with strong acyclicity. In particular, he showed that if we further strengthen the acyclicity requirement to demand an ordinal potential, then this is attained if and only if the game form is dictatorial, i.e., there is at most one voter that can affect the outcome.  He further characterized game forms that are strongly acyclic under \emph{coalitional improvements}, and provided broad classes of game forms that are ``almost unrestricted acyclic,'' i.e. restricted-acyclic under mild restrictions on voters' actions.  
Other partial characterizations have been provided for acyclicity in complete information extensive-form games~\cite{boros2008nash,andersson2010acyclicity}. Some of this work is explained in more detail in the following sections.  

The study of classes of games (i.e. game forms with utilities) that are guaranteed to be acyclic or weakly acyclic attracted much attention, in particular regarding the existence and properties of potential functions~\cite{monderer1996potential,milchtaich1996congestion,fabrikant2010structure,apt2012classification}. 

\paragraph{Strategic voting}

The notion of strategic voting has been highlighted in research on Social Choice as crucial to understanding the relationship between preferences of a population and the final outcome of elections. In various applications (ranging from political domains to artificial intelligence [AI]), the most widely used voting rule is Plurality, in which each voter has one vote and the winner is the candidate who received the highest number of votes. While it is known that no reasonable voting rule is completely immune to strategic behavior~\cite{Gib73,Sat75}, Plurality has been shown to be particularly susceptible, both in theory~\cite{saari:90,friedgut2011quantitative} and in practice~\cite{forsythe:96}. This makes the analysis of any election campaign---even one where the simple Plurality rule is used---a challenging task. As voters may speculate and counter-speculate, it would be beneficial to have formal tools that would help us understand (and perhaps predict) the final outcome. 

In particular, natural tools for this task include the well-studied solution concepts developed for normal form games, such as better/best responses, dominant strategies or different variants of equilibrium. Now, while voting settings are not commonly presented in this way, several natural formulations have been proposed in the past~\cite{dhillon04,Chopra:04,SerSan04,FMT12,messner02}. These formulations are extremely simple for Plurality voting games, where voters have only a few available ways to vote. Specifically, some of this previous work has been devoted to the analysis of solution concepts such as \emph{elimination of dominated strategies}~\cite{dhillon04} and \emph{strong equilibria}~\cite{SerSan04}. 
There has been other multi-step voting procedures that have been proposed in the literature, such as iterated majority vote~\cite{Airiau:09} and extensive form games where voters vote one by one~\cite{desmedt2010equilibria}. In contrast to iterative voting, these models are inconsistent with the better-reply dynamics in normal form games, and are analyzed via different techniques. A model more similar to ours was recently studied in \cite{elkind2015gibbard}, where voters can choose between voting truthfully, and manipulating under the assumption that everyone else are truthful.

Convergence of better-reply dynamics in iterative voting for particular voting rules has been studied extensively in the computational social choice literature. We summarize and compare these findings with ours in the concluding section, and in particular in Table~\ref{tab:results_rules}. 

An important question in every model of strategic voting, including iterative voting, is whether the reached equilibrium is good for the society according to various metrics. Branzei et al.\shortcite{branzeibad} showed bounds on the \emph{dynamic price of anarchy}, i.e. how far can the final outcome be from the initial truthful outcome. Other work used simulations to show that iterative voting may improve the social welfare or Condorcet efficiency~\cite{grandi2013restricted,MLR14,koolyk2016convergence}, but typically under the assumptions that voters use various heuristics.

\paragraph{Biased and sophisticated voting} 
Some recent work on iterative voting deals with voters who are uncertain, truth-biased, lazy-biased, bounded-rational, non-myopic, or apply some other restrictions and/or heuristics that diverge from the standard notion of better-reply in games~\cite{RE12,gohar2012manipulative,grandi2013restricted,obraztsova2013plurality,MLR14,rabinovich2015analysis,obraztsova2015convergence,Meir15}. Although the framework is suitable for studying such iterative dynamics as well, this paper deals exclusively with myopic better-reply dynamics.\footnote{We do consider however two standard ways to handle ties that slightly relax the better-reply definition. See Section~\ref{subsec:SD}.} 

\subsection{Contribution and structure}

The paper unfolds as follows. In Section~\ref{sec:preliminaries}, we define the iterative voting model within the more general framework of game forms and acyclicity properties.
In Section~\ref{sec:FIP} we consider strong acyclicity, and settle an open question regarding the existence of acyclic non-separable game forms. 
Section~\ref{sec:Plurality} focuses on order-free acyclicity of the Plurality rule. Our main result in this section shows that to guarantee convergence, it is necessary and sufficient that voters restrict their actions in a natural way that we term \emph{direct reply}---meaning that a voter will only reassign his vote to a candidate that will become a winner as a result.
In Section~\ref{sec:weak}, we use variations of Plurality to show a strict separation between restricted acyclicity and weak acyclicity, thereby settling another open question. 
 We conclude in Section~\ref{sec:conclusion}.
%


\ssection{Preliminaries}\label{sec:preliminaries}
We usually denote sets by uppercase letters (e.g., $A,B,\ldots$), and vectors by bold letters (e.g., $\vec a=(a_1,\ldots,a_n)$). 


\subsection{Voting rules and game forms}\label{subsec:model}


There is a set $C$ of $m$ alternatives (or \emph{candidates}), and a set $N$ of $n$ strategic agents, or \emph{voters}. A game form  (also called a \emph{voting rule}) $f$ allows each agent $i\in N$ to select an action $a_i$ from a set of messages $A_i$. Thus the input to $f$ is a vector $\vec a =(a_1,\ldots,a_n)$ called an \emph{action profile}. We also refer to $a_i$ as the \emph{vote} of agent $i$ in profile $\vec a$. 
Then, $f$ chooses a winning alternative---i.e., it is a function $f:\calA \rightarrow C$, where $\calA = \times_{i\in N}A_i$. See Fig.~\ref{fig:GF} for examples.  

A voting rule $f$ is \emph{standard} if $A_i=A$ for all $i$, and $A$ is either $\pi(C)$ (the set of permutations over $C$) or a coarsening of $\pi(C)$. Thus most common voting rules except Approval are standard. Mixed strategies are not allowed. The definitions in this section apply to all voting rules unless stated otherwise. For a permutation $P\in \pi(C)$, We denote by $top(P)$ the first element in $P$. 


\paragraph{Plurality}
In the Plurality voting rule we have that $A=C$, and the winner is the candidate with the most votes. 
We allow for a broader set of ``Plurality game forms'' by considering both weighted and fixed voters, and varying the tie-breaking method.
Each of the strategic voters $i\in N$ has an integer weight $w_i\in \mathbb{N}$.
In addition, there are $\hat n$ ``fixed voters'' who do not play strategically or change their vote. The vector $\hatv s\in \mathbb N^m$  (called ``initial score vector'') specifies the number of fixed votes for each candidate.
Weights and initial scores are part of the game form.\footnote{All of our results still hold if there are no fixed voters, but allowing fixed voters enables the introduction of simpler examples, and facilitates some of the proofs, see Remark~\ref{rem:initial}. For further discussion on fixed voters see \cite{elkind2015gibbard}.} 

\begin{figure}[t]
$$
\begin{array}{c|ccc}
 f_1 & a & b & c\\
	\hline
a & a & a & a \\
b & b & b & b \\
c & c & c & c\\
\end{array}
~~~~~ 
\begin{array}{c|ccc}
 f_2 & a & b & c\\
	\hline
a & a & a & a \\
b & a & b & b \\
c & a & b & c\\
\end{array}
~~~~~
\begin{array}{c|cc}
 f_3 & x & y \\
	\hline
a & a & b  \\
b & b & c  \\
c & c & a \\
\end{array}
~~~~~
\begin{array}{c|cccc}
 f_4 & x & y & z & w\\
	\hline
a & ax & ay & az & aw  \\
b & bx & by & bz &  bw \\
c & cx & cy & cz &  cw\\
\end{array}
$$
\caption{\label{fig:GF} Four examples of game forms with two agents. $f_1$ is a dictatorial game form with 3 candidates (the row agent is the dictator). $f_2$ is the Plurality voting rule with 3 candidates and lexicographic tie-breaking. $f_3$ and $f_4$ are non-standard game forms. In $f_3$, $A_1=C=\{a,b,c\}, A_2=\{x,y\}$.  Note that $f_4$ is completely general (there are $3\times 4$ possible outcomes in $C$, one for each voting profile) and can represent any 3-by-4 game.} 
\end{figure}

%
%
%
%

The \emph{final score} of $c$ for a given profile $\vec a \in A^n$  in the Plurality game form $f_{\vec w,\hatv s}$ is the total weight of voters that vote $c$. We denote the final score vector by $\vec s_{\hatv s,\vec w,\vec a}$ (often just $\vec s_{\vec a}$ or $\vec s$ when the other parameters are clear from the context), where $s(c) = \hat {s}(c) + \sum_{i\in N : a_i=c}w_i$. 

Thus the Plurality rule selects some candidate from $W=\argmax_{c\in C} s_{\hatv s,\vec w,\vec a}(c)$, breaking ties according to some specified method. The two primary variations we consider are $f^{PL}_{\hatv s,\vec w}$ which breaks ties lexicographically, and $f^{PR}_{\hatv s,\vec w}$ which selects a winner from $W$ uniformly at random. 
As with $\vec s$, we omit the scripts $\vec w$ and $\hatv s$ when they are clear from the context.


\jr{

For illustration, consider an example in Fig.~\ref{tab:gf}, demonstrating a specific weighted Plurality game form with two agents.
\begin{figure}[t]
\begin{center}

\begin{tabular}{||l||c|c|c||}
\hline
 $f^{PL}_{\vec w,\hatv s}$\!\!  & $a$ &$b$ &$c$ \\
\hline\hline
$a$& {$(14,9,3)$}  $\{a\}$ \!\!& {$(10,13,3)\  \{b\}$} \!\!&  {$(10,9,7)\  \{a\}$ } \!\!\\
\hline
$b$& {$(11,12,3)$}  $\{b\}$ \!\!&  {$(7,16,3)$}  $\{b\}$ \!\!& {$(7,12,7)\  \{b\}$ }\!\!\\
\hline
$c$ & {$(11,9,6)$}  $\{a\}$ \!\!&  {$(7,13,6)\  \{b\}$} \!\!&   {$(7,9,10)\  \{c\}$}\!\! \\
\hline
\end{tabular}
\caption{\label{tab:gf} A game form $f^{PL}_{\vec w,\hatv s}$, where $N=\{1,2\}$, $A_1=A_2=C=\{a,b,c\}$, $\hatv{s}=(7,9,3)$ and $\vec w=(3,4)$ (i.e., voter~1 has weight~3 and voter~2 has weight~4).  The table shows the final score vector~~$\vec s_{(a_1,a_2)}$ for every joint action of the two voters, and the respective winning candidate $f^{PL}_{\vec w,\hatv s}(a_1,a_2)$ in curly brackets. 
}
\end{center}
\end{figure}
}

\subsection{Incentives}\label{subsec:incentives}
Games are attained by adding either cardinal or ordinal utility to a game form. The linear order relation $Q_i \in \pi(C)$ reflects the preferences of agent $i$. That is, $i$ prefers $c$ over $c'$ (denoted $c \succ_i c'$) if $(c,c')\in Q_i$. 
The vector containing the preferences of all $n$ agents is called a \emph{preference profile}, and is denoted by $\vec Q = (Q_1,\ldots,Q_n)$. The game form $f$, coupled with a preference profile $\vec Q$, defines an ordinal utility normal form game $G=\tup{f,\vec Q}$ with $n$ agents, where agent $i$ prefers outcome $f(\vec a)$ over outcome $f(\vec a')$ if $f(\vec a) \ord i f(\vec a')$. In standard game forms the action $a_i$ may indicate the agent's preferences, hence their common identification with voting rules.


\paragraph{Improvement steps and equilibria}\label{subsec:NE}
Having defined a normal form game, we can now apply standard solution concepts. Let $G=\tup{f,\vec Q}$ be a game, and let $\vec a = (\vec a_{-i},a_i)$ be a joint action in $G$. 

We denote by $\vec a \step{i} \vec a'$ an \emph{individual improvement step}, if (1) $\vec a,\vec a'$ differ only by the action of player $i$; and (2) $f(a_{-i},a_i') \ord i f(a_{-i},a_i)$.
We sometimes omit the actions of the other voters $\vec a_{-i}$ when they are clear from the context, only writing $a_i \step{i} a'_i$. 
We denote by $I_i(\vec a)\subseteq A_i$ the set of actions $a'_i$ s.t. $a_i \step{i} a_i'$ is an improvement step of agent $i$ in $\vec a$, and $I(\vec a)=\bigcup_{i\in N}\bigcup_{a'_i\in I_i(\vec a)}(\vec a_{-i},a'_i)$.  $\vec a \step{i} a'_i$ is called a \emph{best reply} if $a'_i$ is $i$'s most preferred candidate in $I_i(\vec a)$. 

A joint action $\vec a$ is a (pure) \emph{Nash equilibrium} (NE) in $G$ if $I(\vec a)=\emptyset$. That is, no agent can gain by changing his vote, provided that others keep their strategies unchanged. A priori, a game with pure strategies does not have to admit any NE. 

Now, observe that when $f$ is a standard voting rule the preference profile $\vec Q$ induces a special joint action $\vec a^*=\vec a^*(\vec Q)$, termed the \emph{truthful state}, where $a^*_i$ equals (the coarsening of) $Q_i$. E.g. in Plurality $a^*_i=top(Q_i)$. We  refer to $f(\vec a^*)$ as the \emph{truthful outcome} of the game $\tup{f,\vec Q}$. 

\jr{
The truthful state may or may not be included in the NE points of the game, as can be seen from Tables~\ref{tab:g1} and~\ref{tab:g2} that demonstrate games that are induced by adding incentives to the game form shown  in Fig.~\ref{tab:gf}, and indicate the truthful states and the NE points in these games.
\begin{figure}[t]
\begin{center}
\begin{tabular}{||l||c|c|c||}
\hline
$\tup{f,\vec Q^1}$  & a &b &* c \\
\hline\hline
* a& $\mathbf{\{a\}\ 3,2 }$ & $\{b\}\ 2,1$ & * $ \mathbf{\{a\}\ 3,2}$  \\
\hline
b& $\{b\}\ 2,1$ &  $ \mathbf{\{b\}\ 2,1}$ &  $\{b\}\  2,1$ \\
\hline
c& $\{a\}\ 3,2$ &  $\{b\}\ 2,1$ &   $\{c\}\ 1,3$\\
\hline
\end{tabular}
\caption{\label{tab:g1} A game $G=\tup{f,\vec Q^1}$, where $f=f^{PL}_{\vec w,\hatv s}$ is as in Fig.~\ref{tab:gf}, and $\vec Q^1$ is defined by $a \ord 1 b\ord 1 c$ and $c\ord 2 a\ord 2 b$. The table shows the ordinal utility of the outcome to each agent, where $3$ means the best candidate. \textbf{Bold} outcomes are the NE points. Here the truthful vote (marked with *) is also a NE.
}
\end{center}
\end{figure}

\begin{figure}[t]
\begin{center}
\begin{tabular}{||l||c|c|c||}
\hline
$\tup{f,\vec Q^2}$  & $a$ &$b$ &* $c$ \\
\hline\hline
* $a$& $\{a\}\ 3,1 $ & $ \mathbf{\{b\}\ 1,2}$ & * $\{a\}\ 3,1$  \\
\hline
 $b$& $\{b\}\ 1,2$ &  $ \mathbf{\{b\}\ 1,2}$ &  $\{b\}\  1,2$ \\
\hline
$c$ & $\{a\}\ 3,1$ &  $\{b\}\ 1,2$ &   $\{c\}\ 2,3$\\
\hline
\end{tabular}
\caption{\label{tab:g2} This game has the same game form as in Fig.~\ref{tab:gf}, and the preference profile $\vec Q^2$ is $a \ord 1 c\ord 1 b$ and $c\ord 2 b\ord 2 a$. In this case, the truthful vote $\vec a^*(\vec Q^2)$ is not a NE. \vspace{-8mm}}
\end{center}
\end{figure}
}

\subsection{Iterative Games}\label{subsec:dynamics}

We consider natural \emph{dynamics} in iterative games. Assume that agents start by announcing some initial profile $\vec a^0$, and then proceed as follows: at each step $t$ a single agent $i$ may change his vote to $a'_i\in I_i(\vec a^{t-1})$, resulting in a new state (joint action) $\vec a^t= (\vec a^{t-1}_{-i},a'_i)$. The process ends when no agent has objections, and the outcome is set by the last state. 

\newpar{Local improvement graphs and schedulers}
Any game $G$ induces a directed graph whose vertices are all action profiles (states) $\calA$, and edges are all local improvement steps~\cite{young1993evolution,andersson2010acyclicity}. The pure Nash equilibria of $G$ are all states with no outgoing edges. 
Since a state may have multiple outgoing edges ($|I(\vec a)|>1$), we need to specify which one is selected in a given play. 
	
	A \emph{scheduler} $\phi$ selects which edge is followed at state $\vec a$ at any step of the game~\cite{apt2012classification}. The scheduler can be decomposed into two parts, namely selecting an agent $i$ to play (agent scheduler $\phi^N$), and selecting an action in $I_i(\vec a)$ (action scheduler $\phi^A$), where $\phi=(\phi^N,\phi^A)$. We note that a scheduler may or may not depend on the history or other factors, but this does not affect any of our results.

	
	\newpar{Convergence and acyclicity}
	Given a game $G$, an initial action profile $\vec a^0$ and a scheduler $\phi$, we get a unique (possibly infinite) path of steps.\footnote{By ``step'' we mean an individual improvement step, unless specified otherwise.} Also, it is immediate to see that the path is finite if and only if it reaches a Nash equilibrium (which is the last state in the path).   
	We say that the triple $\tup{G,\vec a^0,\phi}$ \emph{converges} if the induced path is finite.
	
  %
	Following \cite{monderer1996potential,milchtaich1996congestion}, a game $G$ has the \emph{finite individual improvement property} (we say that $G$ \emph{is} FIP),  if $\tup{G,\vec a^0,\phi}$ converges for \emph{any} $\vec a^0$ and scheduler $\phi$. Games that are  FIP  are also known as \emph{acyclic games} and as \emph{generalized ordinal potential games}~\cite{monderer1996potential}. 
	
	It is quite easy to see that not all Plurality games are FIP (see examples in Section~\ref{sec:Plurality}).
	%
	 However, there are alternative, weaker notions of acyclicity and convergence.
	
	\begin{itemize}[itemsep=1mm]
		\item 	A game $G$ is \emph{weakly-FIP}  if there is \emph{some} scheduler $\phi$ such that $\tup{G,\vec a^0,\phi}$ converges for any $\vec a^0$. Such games are known as \emph{weakly acyclic}, or as $\phi$-potential games~\cite{apt2012classification}.
\item A game $G$ is \emph{restricted-FIP} if there is \emph{some action scheduler} $\phi^A$ such that $\tup{G,\vec a^0,(\phi^N,\phi^A)}$ converges for any $\vec a^0$ and $\phi^N$~\cite{kukushkin2011acyclicity}. We term such games as \emph{order-free acyclic}.
	\end{itemize}
	Intuitively, restricted FIP means that there is some restriction players can adopt s.t. convergence is guaranteed regardless of the order in which they play. Kukushkin identifies a particular restriction of interest, namely restriction to best-reply improvements, and defines the \emph{finite best-reply property} (FBRP) and its weak and restricted analogs. 
	We emphasize that an action scheduler \emph{must} select an action in $I_i(\vec a)$, if one exists. Thus restricted dynamics that may disallow all available actions (as in \cite{gohar2012manipulative,grandi2013restricted}) do not fall under the definition of restricted-FIP (but can be considered as separate dynamics).  
	
	We identify a different restriction, namely \emph{direct reply}, that is well defined under the Plurality rule. Formally, a step $\vec a \step i \vec  a'$ is a direct reply if $f(\vec a')=a'_i$, i.e., if  $i$ votes for the new winner (see labeled examples in Section~\ref{sec:Plurality}).
	Another rule where a natural direct reply exists is Veto, where a voter can veto the current winner~\cite{lev2012convergence}.
	
	\rmr{not here: While there is also a natural direct reply under Veto (to veto the current winner), it is not obvious how to define direct replies in other voting rules. A good candidate strategy for positional scoring rule is to rank the new winner first, and then all other candidates by increasing order of their current score. Note that the direct replies for Plurality and Veto are special cases.  }
	$\phi^A$ is direct if it always selects a direct reply. We get the  following definitions for a Plurality game $G$, where FDRP stands for \emph{finite direct reply property}:
	\begin{itemize}[itemsep=1mm]
\item  $G$ is \emph{FDRP} if $\tup{G,\vec a^0,\phi}$ converges for any $\vec a^0$ and any direct  $\phi$.
		\item 	 $G$ is \emph{weakly-FDRP} if there is a  direct   $\phi$ such that $\tup{G,\vec a^0,\phi}$ converges for any $\vec a^0$.
		\item  $G$ is \emph{restricted-FDRP} if there is a direct  $\phi^A$ such that $\tup{G,\vec a^0,(\phi^N,\phi^A)}$ converges for any $\vec a^0$ and $\phi^N$.
		\item FDBRP means that replies are both best and direct. Note that it is unique and thus cannot be further restricted.
	\end{itemize}
Finally, a game form $f$ has the X property (where X is any of the above versions of finite improvement) if $\tup{f,\vec Q}$ is X for all preference profiles $\vec Q\in(\pi(C))^n$.
	We have the following entailments, both for games and for game forms. 
	The third row is only relevant for Plurality/Veto. 
	$$\begin{tabular}{ccc|ccc|ccccc}
  &&FBRP &&& & restricted-FBRP               & $\Rightarrow$ &  weak-FBRP &&\\
	  &&    $\Uparrow$ & &  &&         $\Downarrow$      & &   $\Downarrow$      &&\\
	   ordinal potential  &$\Rightarrow$&   FIP & $\Rrightarrow$& FDBRP& $\Rrightarrow$ &          restricted-FIP    & $\Rightarrow$ & weak-FIP &$\Rightarrow$ & pure Nash \\		
					          exists &&    $\Downarrow$      & & & & $\Uparrow$ & &   $\Uparrow$ & & exists \\
				&&FDRP && &&restricted-FDRP               & $\Rightarrow$ &  weak-FDRP &&\\
	       \end{tabular}$$

 Kukushkin notes that there are no known examples of game forms that are weak-FIP, but not restricted-FIP. We settle this question later in Section~\ref{sec:weak_weighted}.


\newpar{Convergence from the truth}
We say that a game $G$ is \emph{FIP from state $\vec a$} if $\tup{G,\vec a,\phi}$ converges for any $\phi$.  Clearly a game is FIP iff it is FIP from $\vec a$ for any $\vec a\in A^n$. The definitions for other all other notions of finite improvement properties are analogous. 

We are particularly interested in convergence from the truthful state $\vec a^*$. This is since: a. it is rather plausible to assume that agents will start by voting truthfully, especially when not sure about others' preferences; and b. even with complete information, they may be inclined to start truthfully, as they can always later change their vote. 

\jr{
\paragraph{Heuristic voting}
Much work on iterative voting deals with heuristics, rather than best- or better-replies. Strong, Restricted, and Weak convergence properties can be defined the same way, where the only difference is the way we define $I_i(\vec a)$ (i.e., all steps that are allowed for agent $i$ at state $\vec a$ by the considered heuristics). For example, \emph{truth-bias} assumes that if a voter does not have any local improvement step, she reverts to her truthful vote~\cite{obraztsova2013plurality}.
 Some heuristics are already restricted to a single action (for example, ``k-pragmatist''~\cite{grandi2013restricted}). In these cases the only meaningful distinction is between FIP and weak-FIP. 
In this paper we do not consider heuristic voting. 
}



\ssection{Strong Acyclicity}\label{sec:FIP}
An \emph{ordinal potential} is a function that strictly increases if and only if some agent plays a better-reply~\cite{monderer1996potential}. A \emph{generalized ordinal potential} is a function that strictly increases with every better-reply, but may also increase with other steps. Clearly, a game is FIP if and only if it has a generalized potential  (by a topological sort of the better-reply graph).  

\begin{theorem}[Kukushkin~\shortcite{kukushkin2011acyclicity}]
A game form $f$ guarantees an ordinal potential (i.e. every derived game has an ordinal potential function) if and only if $f$ is a dictatorship.
\end{theorem}
  We emphasize that this result \emph{does not} preclude the existence of other game forms with FIP (\emph{generalized} ordinal potential). Indeed, Kukushkin provides a partial characterization of FIP game forms. For example, a rule where there is a linear order $L$ over $C$, and the winner is the first candidate according to $L$ that is top-ranked by at least one voter.  

A game form $f$ is called ``separable''~\cite{kukushkin2011acyclicity} if there are mappings $g_i:A_i\rightarrow C$ for $i\in N$ s.t. for all $\vec a \in\calA$, $f(\vec a)\in\{g_1(a_1),g_2(a_2),\ldots, g_n(a_n)\}$. That is, the vote of each voter is mapped to a single candidate via some function $g_i$, and the outcome is always one of the candidates in the range. Examples of separable rules include Plurality and dictatorial rules, in both of which $g_i$ are the identity functions.

\begin{conjecture}[Kukushkin~\shortcite{kukushkin2011acyclicity}] Any FIP game form is separable.
\end{conjecture}
Some weaker variations of this conjecture have been proved. In particular, for game forms  with  finite \emph{coalitional improvement} property~\cite{kukushkin2011acyclicity}, and for FIP game forms with $n=2$ voters~\cite{boros2010acyclic} (separable game forms are called ``assignable'' there). 
We next show that for sufficiently large $n$, there are non-separable FIP game forms, thereby refuting the conjecture. Our proof uses the probabilistic method: we sample a game form from some space, and prove that with positive probability it must be non-separable and FIP.

\begin{theorem}\label{th:FIP}
For any $n\geq 20$, there is a non-separable game form $f_n$ s.t. $f_n$ is FIP.
\end{theorem}
\begin{proof}
Let $C=\{a^1,\ldots,a^{2n}\} \cup \{z\}$. 
Let $A_i = \{x,y\}$ for each voter. Thus $f_n$ is a function from the $n$ dimensional binary cube $\calB=\{x,y\}^n$ to $C$.
We select $2n$ profiles $\vec a^1,\ldots,\vec a^{2n}$ uniformly at random, i.i.d. from $\calB$ (allowing repetitions), and define $f_n(\vec a^j)=a^j$ for all $j\in \{1,\ldots,2n\}$. For all other $2^n-2n$ profiles we define $f_n(\vec a)=z$.


For any two profiles $\vec a,\vec a'$, let $d(\vec a,\vec a')$ be the number of voters that disagree in $\vec a,\vec a'$ (the Manhattan distance on the cube). 
Let $B\subseteq \calB$ be all $2n$ profiles whose outcome is not $z$. For $j,j'\leq 2n$, denote by $p_{j,j'}$ the probability that  $d(\vec a^j,\vec a^{j'})\leq 2$, and by $X_{j,j'}$ the corresponding indicator random variable. Since both of $\vec a^j,\vec a^{j'}$ were sampled uniformly i.i.d., and there are less than $n^2$ profiles within distance $2$ from $\vec a^j$, we get that $p_{j,j'} \leq \frac{n^2}{2^n}$. 

Next, by the union bound, 
$$Pr(\exists X_{j,j'}=1) \leq \sum_{j\leq 2n,j'\leq 2n}Pr(X_{j,j'}) = \sum_{j\leq 2n,j'\leq 2n}p_{j,j'}\leq (2n)^2\frac{n^2}{2^n}=\frac{4n^2}{2^n},$$
 which is strictly less than $1$ for $n\geq 20$. Thus w.p. $>0$ we get $X_{j,j'}=0$ for all $j,j'$. In particular there is at least one such game form $f^*_n$ where $X_{j,j'}=0$ for all $j,j'$. We argue that $f^*_n$ is both  FIP and non-separable.\footnote{Using the Hamming error-correcting code~\cite{hamming1950error}, it is in fact possible to explicitly construct $f^*_n$ for as few as $n=7$ voters. The rest of the proof remains the same.}

Assume towards a contradiction that there is some cycle of better-replies in $f^*_n$. Then there must be a path containing at least $3$ distinct outcomes, and thus at least $2$ profiles from $B$. Denote these profiles by $\vec a,\vec b$. Since $X_{j,j'}=0$ for all $j,j'$, we have that any path between $\vec a$ and $\vec b$ is of length at least $3$, and that the path must contain at least two consequent states whose outcome is $z$. This path cannot be a better-reply path, since a better reply must change the outcome. Hence we get a contradiction and $f^*_n$ is FIP. 

Finally, note that since $X_{j,j'}=0$ for all $j,j'$, in particular $\vec a^j$ are all distinct profiles, and thus $f^*_n$ has $2n+1>\sum_{i\leq n}|A_i|$ possible outcomes.
In contrast, for any separable rule $f$ the size of the range of $f$ is at most $\sum_{i\leq n}|A_i|$, since $f(\vec a)=g_i(b)$ for some $i\in N$ and $b\in A_i$. This means that $f^*_n$ is non-separable. 
\end{proof}

For most common voting rules, separable or not, it is easy to find examples where some cycles occur.  Thus one should focus on the weaker notions of convergence discussed in Section~\ref{sec:intro}, which is what we do in the remainder of the paper.


\ssection{Order-Free Acyclicity: Plurality}\label{sec:Plurality}

\jr{

\paragraph{Improvement steps in Plurality}
Recall that along a given path, $\vec a^t\in A^n=C^n$ denotes the voting profile at time $t$. 
We denote by $\vec s^t = \vec s_{\vec a^t}$ the score vector at time $t$; by $cw^t = f^{PL}(\vec a^t)$ the candidate that wins at time $t$; and by $sw^t = \ddot s^t(cw^t)$ the highest score at time $t$ (including tie-breaking if it applies).

Suppose that agent $i$ has an improvement step (a.k.a. \emph{better reply}) $a_i^{t-1} \step{i} a_i^{t} $ at time $t$. 
We classify all possible steps into the following types (an example of such a step appears in parentheses):

\begin{enumerate}[label*=Type~\arabic*.]
	\item from $a_{i}^{t-1} \neq cw^{t-1}$ to $a_{i}^{t} =cw^t$~; (step~1 in Ex.\ref{ex:det_better}a.)
	\item from $a_{i}^{t-1} = cw^{t-1}$ to $a_{i}^{t} = cw^t$~; (step~1 in Ex.\ref{ex:det_better}b.),
	\item from $a_{i}^{t-1} = cw^{t-1}$ to $a_{i}^{t}\neq cw^t$~; (step~2 in Ex.\ref{ex:det_better}a.)
\end{enumerate}
Note that steps of type 1 and 2 are direct, whereas type 3 steps are indirect. 
}
%

 \subsection{Lexicographic Tie-Breaking}\label{sec:lexicographic}

In this section we assume that ties are broken lexicographically.
Given some score vector $\vec s$, we denote by $\ddot{s}(c)\in \mathbb R$ the score of $c \in C$ that includes the lexicographic tie-breaking component. One way to formally define it is by setting $\ddot{s}(c) = s(c) + \frac{1}{m+1}(m-L(c))$, where $L(c)$ is the lexicographic index of candidate $c$. However the only important property of $\ddot{\vec s}$ is that $\ddot s(c) > \ddot s(c')$ if either $s(c)>s(c')$ or the score is equal and $c$ has a higher priority (lower index) than $c'$. 

Thus for Plurality with lexicographic tie-breaking, a given weight vector $\vec w$ and a given initial score vector $\hatv s$, we denote the outcome by
$$f^{PL}_{\hatv s,\vec w}(\vec a) = \argmax_{c\in C}\ddot s_{\hatv s,\vec w,\vec a}(c).$$

As with $\vec s$, we omit the scripts $\vec w,\hatv s$ and $PL$ when they are clear from the context.

\begin{lemma}
Consider a game $\tup{f^{PL}_{\vec w,\hatv s},\vec Q}$.
If there exists a better reply for a given agent $i$ at state $\vec a^{t-1}$, then $i$ has a direct best reply at state $\vec a^{t-1}$. 
\end{lemma}
The proof is trivial under lexicographic tie-breaking, by letting $i$ vote for her most preferred candidate among all better replies. In this case the direct best reply is also unique. 

One implication of the lemma is that it is justified and natural to restrict our discussion to direct replies and focus on FDRP, as w.l.o.g.\ a voter always has a direct reply that is at least as good as any other reply.
%


\paragraph{Unweighted Voters}\label{subsec:unweighted}
Suppose all voters have unit weight. We start with our main result for this section. 
\begin{theorem} 
\label{th:FDRP} 
$f^{PL}_{\hatv s}$ is FDRP. Moreover, any path of direct replies will converge after at most $m^2 n^2$ steps. In particular, Plurality is order-free acyclic.
\end{theorem}

This extends a weaker version of the theorem that appeared in the preliminary version of this paper~\cite{MPRJ:2010:AAAI}, which only showed FDBRP. 
The bound on the number of direct-best-reply steps was recently improved to $O(mn)$ in \cite[Theorem~5.4]{reyhani2012best}.

\jr{
\begin{proof}
By our restriction to direct replies, there can only be moves of types~1 and 2. We first consider moves of type~1, and inductively prove two invariants that yield a bound on the total number of such moves. Next, we bound the number of moves of type~2 by a given voter between any of his moves of type~1, which completes the proof. 

Consider time $t-1$ and denote the score of the current winner (including tie-breaking) by $\bar s =  sw^{t-1}$. Suppose that a move $a \step{i} b$ of type~1 occurs at time $t$: that is, $a \neq cw^{t-1}$ and $b =cw^t$.  We then have (see Fig.~\ref{fig:proof1}): 
\labeq{step_t}{
\ddot s^t(b) = sw^t \geq sw^{t-1} = \bar s \geq \ddot s^{t-1}(a) = \ddot s^t(a)+1.}

\input{example_proof_DFRP}
We claim that at any later time $t'\geq t$ the following two invariants hold: 
\begin{enumerate}[label*=\Roman*.]
	\item  Either there is a candidate $c\neq a$ whose score is at least $\bar s+1$, or there are at least two candidates $c,c'\neq a$ whose score is at least $\bar s$. In particular it holds in either case that $sw^{t'}\geq \bar s$. 
	\item  The score of $a$ does not increase: $\ddot s^{t'}(a) \leq \ddot s^t(a)$.
\end{enumerate}
Note that this, coupled with Eq.~\eqref{eq:step_t}, implies that candidate $a$ will never win again, as its score will stay strictly below $\bar s$, and there will always be a candidate with a score of at least $\bar s$.  

We now prove both invariants by induction on time $t'$. In the base case $t'=t$, (I) holds since both $cw^{t-1}$ and $b$ have a score of at least $\bar s$, and (II) holds trivially.

Assume by induction that both invariants hold until time $t'-1$, and consider step $t'$ by voter $j$.
Due to (I), we either have at least two candidates whose score is at least $\bar s$, or a candidate with a score of at least $\bar s+1$. Due to (II) and Eq.~\eqref{eq:step_t} we have that $\ddot s_{t'}(a) \leq \ddot s_t(a) < \bar s-1$. 

Let $d \step{j}  d'$ be the step at time $t'$ by voter $j$ (that is, $d=a_{j}^{t'-1}, d'=a_j^{t'}$). We first argue that $d'\neq a$: by adding the vote of $j$ to $a$ its score will still be strictly less than $\bar s$, whereas by removing a vote from any other candidate $d$, we still have at least one candidate $c$ with score at least $\bar s$. Thus $a$ cannot be a direct reply for any voter $j$, and (II) still holds after step $t'$. 

It remains to show that (I) holds. 
If $d$ is \emph{not} one of the candidates in (I) with the score of at least $\bar s$ at time $t'-1$, then their score does not decrease after step $t'$, and we are done. Otherwise, we divide into the following cases:

\begin{enumerate}
	\item At $t'-1$, $d$ is the (only) candidate with a score of at least $\bar s +1$.
	\item At $t'-1$, candidates $c,c'$ have scores of at least $\bar s$, and $d$ is one of them (w.l.o.g.\ $d=c$).	
\end{enumerate}

In the first case, $\ddot s^{t'}(d) = \ddot s^{t-1}(d)-1 \geq \bar s+1-1 = \bar s$, whereas $\ddot s^{t'}(d') > \ddot s^{t'}(d) \geq \bar s$.
Thus both $d,d'$ have scores of at least $\bar s$ at time $t'$, as required. In the second case, since only $c=d$ can lose votes, then if $d'\neq c'$,
$$\ddot s^{t'}(d') = sw^{t'} \geq  \ddot s^{t'}(c') = \ddot s^{t'-1}(c') \geq \bar s,$$
and thus both $c',d'$ have scores of at least $\bar s$ at time $t$, as required. If $d'=c'$, then 
and thus both $c',d'$ have scores of at least $\bar s$ at time $t$, as required. If $d'=c'$, then 
$$\ddot s^{t'}(d') = \ddot s^{t'-1}(d')+1 =  \ddot s^{t'-1}(c')+1 \geq \bar s+1,$$
that is, $d'$ has a score of at least $\bar s+1$, as required. 

Next, we demonstrate that invariants (I) and (II) supply us with a polynomial bound on the rate of convergence. Indeed, as we mentioned before, at every step of type~1, at least one candidate is ruled out permanently, and there are at most $n$ times that a vote can be withdrawn from a given candidate. Also note that, since a type~2 move by a given voter~$i$ implies that he prefers $a^t_i$ to $a^{t-1}_i$, each voter can make at most $m-1$ type~2 moves before making a move of type~1.  Hence, there are in total at most $m^2 n^2$ steps until convergence.
\end{proof}

Furthermore, it is easy to  show that if all voters start from the truthful state then type~2 moves never occur.  Thus, the score of the winner never decreases, and the game converges in at most $m n$ steps. 

Next, we show that the restriction to direct replies is necessary to guarantee convergence, whereas a restriction to best replies is insufficient.
}

\begin{proposition} 
\label{ex:det_better} 
\label{ex:no_FBRP}
$f^{PL}$  is not FBRP, even from the truthful state.
 Moreover, there are: (a) a counterexample with two strategic agents and an arbitrary initial state; (b)  a  counterexample with three strategic agents and a truthful initial vote.
\end{proposition}

\begin{remark}
\label{rem:initial}
In this example and in others throughout the paper we use an initial score vector $\hatv s$. However, this is w.l.o.g. since we could replace $\hatv s$ with additional voters that do not participate in the cycle. Initial scores are only useful to construct examples that are simpler and/or with fewer strategic agents. This holds for all negative results in the paper.\footnote{Note that the remark does no longer hold if $\hatv s$ is used to construct a counter example for weak-FIP. However we use no such examples in this paper.} For positive results, we have to show convergence for every initial scores $\hatv s$.
\end{remark}

\begin{xproof}{~\ref{ex:det_better}a}
$C=\{a,b,c\}$.  We have a single fixed voter voting for $a$, thus $\hat{\vec s} = (1,0,0)$. 
The preference profile is defined as $a\ord 1 b \ord 1 c$, $\ c\ord 2 b \ord 2 a$. 
The following cycle consists of better replies (the vector denotes the votes $(a_1,a_2)$ at time $t$, the winner appears in curly brackets):
$$(b,c) \{a\} \step{2} (b,b) \{b\} \step{1} (c,b) \{a\} \step{2} (c,c) \{c\} \step{1} (b,c).$$
Note that all steps are best-replies, but the steps of agent~1 are indirect.
\end{xproof}%

\begin{xproof}{~\ref{ex:det_better}b}
$C=\{a,b,c,d\}$. Candidates $a,b$, and $c$ have 2 fixed voters each, thus $\hatv s = (2,2,2,0)$. We use 3 agents with the following preferences: $d\ord 1 a\ord 1 b \ord 1 c$, $\  c\ord 2 b \ord 2 a \ord 2 d$ and $\  d\ord 3 a\ord 3 b \ord 3 c$. 
Starting from the truthful state $(d,c,d)$ the agents can make the following two improvement steps, which are direct best-replies (showing only the outcome $\vec s$ and the winner):
$(2,2,3,2) \{c\} \step{1} (2,3,3,1) \{b\} \step{3} (3,3,3,0) \{a\}$,\\
after which agents~1 and 2 repeat the cycle shown in~(\ref{ex:det_better}a).
\end{xproof}

\jr{
Thus for the non-weighted lexicographic case Theorem~\ref{th:FDRP} and Proposition~\ref{ex:det_better} provide a clear-cut rule: direct replies guarantee convergence, whereas convergence is not guaranteed  under other restrictions such as best reply or initial truthful vote. However, as the following section demonstrates, in the presence of weighted agents even direct replies may no longer converge.  
}



\paragraph{Weighted Voters}\label{subsec:weighted}
\jr{
Next, we show that if the voters may have non-identical weights, then convergence to equilibrium is not guaranteed even if they start from the truthful state and use direct best replies. 
}

\begin{proposition}
\label{ex:no_RFDRP_w} 
There is $f^{PL}_{\vec w}$ that is not restricted-FDRP, even from the truthful state.
\end{proposition}

\begin{xproof}{\ref{ex:no_RFDRP_w}}
The initial fixed score of candidates $\{a,b,c,d\}$ is $\hatv s = (0,1,2,3)$. The weight of each voter $i\in \{1,2,3\}$ is $i$. The preference profile is as follows: $c \ord 1 d \ord 1 b \ord 1 a$, $ b \ord 2 c \ord 2 a \ord 2 d$,  and $a \ord 3 b \ord 3 c \ord 3 d$. We omit the rest of the proof. 
\jr{
The initial truthful profile is thus $\vec a^0=(c,b,a)$, which results in the score vector $\vec s^0 = (3,3,3,3)$ where $a$ is the winner. 
$$
\begin{array}{lccccc}
\text{votes:} & (c,b,a) & \step{1} & (d,b,a) &\step{2} & (d,c,a) \\
\text{scores:} & (3,3,3,3) \{a\} &          & (3,3,2,4) \{d\}&         &   (3,1,4,4)\{c\}\\     
&  \ustep{3} & & & &\dstep{3}\\ 
        &    (c,b,b)   & \lstep{2} & (c,c,b)  &   \lstep{1} & (d,c,b) \\
				&     (0,6,3,3) \{b\} &          & (0,4,5,3) \{c\} &         & (0,4,4,4) \{b\} \\
\end{array}
$$
Our example shows a cycle of direct responses.
Note that at every step there is only one direct reply available to the agent, thus it is not possible to eliminate the cycle by further restricting the action scheduler.  
}
\end{xproof}

%

\jr{
If there are \emph{only two} weighted voters (and possibly other fixed voters), either restriction to direct reply or to a truthful initial state is sufficient to guarantee convergence.
}

\begin{theorem}
\label{th:FDRP_w2} 
$f^{PL}_{\hatv s,\vec w}$ is FDRP for $n=2$.
\end{theorem}

\begin{proof} 
Clearly, in one of the two first states, the agents vote for distinct candidates. At any later state, they must continue voting for distinct candidates, as every step must change the winner, and the other voter is always voting for the current winner. This means that the score of the winner strictly increases with every step (possibly except the first one).  
\end{proof}

\begin{theorem}
\label{th:FIP_w2_t} 
$f^{PL}_{\hatv s,\vec w}$ is FIP from the truth for $n=2$.
\end{theorem}

\begin{proof}
We show that the score of the winner can only increase. This clearly holds in the first step, which must be of type~1. Once again, we have that the two agents always vote for different candidates, and thus only steps that increase the score can change the identity of the winner.
\end{proof}
Thus in either case convergence is guaranteed after at most $2m$ steps.


It remains an open question whether there is any restriction on better replies that guarantees order-free acyclicity in weighted games, i.e. if $f^{PL}_{\vec w}$ is restricted-FIP for $n>2$. However Prop.~\ref{ex:no_RFDRP_w} shows that if such restricted dynamic exists, it must make use of indirect replies, which is rather unnatural. We thus conjecture that such restricted dynamics does not exist.

%
%
%

\subsection{Arbitrary tie-breaking}\label{subsec:arbitrary}
Lev and Rosenschein~\shortcite{lev2012convergence} showed that for any positional scoring rule (including Plurality), we can assign some (deterministic) tie breaking rule, so that the resulting voting rule may contains cycles. For any positional scoring rule $f_\alpha$ with score vector $\alpha$, denote by $f_\alpha^{LR}$ the same rule with the Lev-Rosenschein tie-breaking. 

\begin{proposition}[Theorem~1 in \cite{lev2012convergence}]
$f_\alpha^{LR}$ is not FBRP for any $\alpha$, even for $n=2$, and even from the truth. In particular, Plurality with the Lev-Rosenschein tie-breaking ($f^{PLR}$) is not FBRP.
\end{proposition}

In fact, a slight modification of their example (switching $a$ and $b$ in voter 2's preferences) yields the following:

\begin{proposition}
$f^{PLR}$ is not restricted-FIP, even for $n=2$, and even from the truth. 
\end{proposition}

%

\subsection{Randomized tie-breaking}
\label{sec:randomized}
\jr{
Compared to the previously considered deterministic rule, randomized tie-breaking has the advantage of being neutral---no specific candidate or voter is preferred over another.
}
Formally, the game form $f_{\hatv s,\vec w}^{PR}$ maps any state $\vec a\in A^n$ to the set $\argmax_{c\in C}s_{\hatv s,\vec w,\vec a}(c)$.
Since under randomized tie-breaking there are multiple winners, let $W^t = f^{PR}(\vec a^t)\subseteq C$ denote the set of winners at time $t$.\footnote{This is a slight abuse of the notation we introduce in the beginning, where we defined the set of possible outcomes of $f$ to be $C$. Here we allow any $W\in 2^C\setminus \{\emptyset\}$ as a possible outcome.}
 We define a direct reply $a^{t-1}_i \step i a^{t}_i$ as one where $a^t_i \in W^t$.

If ties are broken randomly, $\ord i$ does \emph{not} induce a complete order over outcomes. For instance, the order $a\ord i b \ord i c$ does not determine if $i$ will prefer $\{b\}$ over $\{a,c\}$. However, we can naturally extend $Q_i$ to a \emph{partial preference order} over subsets.
There are several standard extensions, using the following axioms:\footnote{We thank an anonymous reviewer for the references.}
\begin{description}
	\item[\textbf{K}] (Kelly~\cite{kelly1977strategy}): (1) 	$(\forall a \in X, b\in W, a \succ_i b)\ \Rightarrow\ X \succ_i W$;  (2) $(\forall a \in X, b\in W, a \succeq_i b)\ \Rightarrow\ X \succeq_i W$; 
	\item[\textbf{G}]	(G\"ardenfors~\cite{gardenfors1976manipulation}): $(\forall b \in W, a \ord i b)\ \Rightarrow\  \{a\} \succ_i (\{a\} \cup W) \succ_i W$;
	\item[\textbf{R}] (Responsiveness~\cite{roth1985college}): $a \ord i b\ \iff\ \forall W\subseteq C\setminus\{a,b\},\ (\{a\} \cup W) \succ_i (\{b\} \cup W)~.$
\end{description}
 The axioms reflect various beliefs a rational voter may have on the tie-breaking procedure: the K axiom reflects no assumptions whatsoever; 
The K+G axioms are consistent with tie-breaking according to a fixed and unknown order~\cite{geist2011automated}; and K+G+R axioms are consistent with random tie-breaking with equal probabilities (see Lemma~\ref{lemma:rel} and Prop.~\ref{th:SD_axioms}). In this section we assume all axioms hold, however our results  do not depend on these interpretations, and we do not specify the voter's preferences in cases not covered by the above axioms. Under strict preferences, it also holds that G entails K~\cite{endriss2013sincerity}. We can also define ``weak'' variants G2 and R2 for axioms G and R, by replacing all strict relations with weak ones, however as long as we restrict attention to strict preferences over elements the weak variants are not required.

For the following lemma we only need Axiom~K, i.e. it does not depend on the voter's tie-breaking assumptions.
\begin{lemma}
If there exists a better-reply in $f_{\hatv s,\vec w}^{PR}$ for agent $i$ at state $\vec a^{t-1}$, then $i$ has a direct best-reply. 
\end{lemma}
\begin{proof}
Suppose there is a better reply $a^{t-1}_i \step{i} b$ at time $t-1$. 
As some best reply always exists, denote  by $b'$ an arbitrary best reply. 
Let $W  = f_{\hatv s,\vec w}^{RP}(\vec a_{-i}^{t-1},b')$, and let $a'$ be the most preferred candidate of $i$ in $W$. 
Then we argue that $a^{t-1}_i \step{i} a'$ is a direct best reply of $i$. Since $a'$ is a direct reply by definition, it is left to show that $a'$ is a best reply 
(for the lexicographic case this follows immediately from $W=\{a'\}$ and $f^{PL}(\vec a_{-i}^{t-1},a') = W = \{a'\}$).

If $b'$ is a direct reply then $b'=a'$ and we are done. Thus assume that $b'$ is not a direct reply from $a^{t-1}_i \step{i}$.
 Then $b'\notin W$. By voting for $a'\in W$, we get that $f_{\hatv s,\vec w}^{RP}(\vec a_{-i}^{t-1},a')=\{a'\}$, i.e., $a'$ remains the unique winner. If $|W|=1$ then we are done as in the lexicographic case. Otherwise we apply Axiom~K2  with $X = \{a'\}$, and get that $a' \succeq_i  W$. 
That is, 
$$f_{\hatv s,\vec w}^{RP}(\vec a_{-i}^{t-1},a') = \{a'\} \succeq_i W = f_{\hatv s,\vec w}^{RP}(\vec a_{-i}^{t-1},b'),$$
which means that $a'$ is also a best-reply.
\end{proof} 

With weighted votes and and random tie-breaking, there may not be any pure Nash equilibrium at all~\cite{MPRJ:2010:AAAI}. We therefore restrict attention in the rest of this section to unweighted votes. 
\jr{
\begin{proposition}
\label{ex:noFIP_r} 
$f^{PR}$  is not FIP.
\end{proposition}
\begin{xproof}{\ref{ex:noFIP_r}}
$C=\{a,b,c\}$ with initial score $\hat{\vec s} = (0,1,0)$. The initial state is $\vec a_0 = (a,a,b)$---that is, $\vec s(\vec a_0) = (2,2,0)$ and the outcome is the winner set $\{a,b\}$. The preferences are $a \ord 1 c \ord 1 b$, $b \ord 2 a \ord 2 c$ and $c \ord 3 b \ord 3 a$. We get the following cyclic sequence:
$$\begin{array}{ccccc}
(2,2,0) \{a,b\} &\step{2} &(1,2,1) \{b\} & \step{1} &(0,2,2) \{b,c\}\\
 \ustep{3} & & & &\dstep{3}\\ 
	 (1,2,1) \{b\}& \lstep{1}&(2,1,1) \{a\}  & \lstep{2}&(1,1,2) \{c\} 
	\end{array}
	$$ 
	We emphasize that each step is justified as a better reply by either Axiom~K or Axiom~G. E.g, in the step of agent~2 in the top row, agent~2 prefers $b\succ_2 a$, and thus $b \succ_2 \{a,b\}$ by Axiom~G. This will be used later in Section~\ref{subsec:SD}.
\end{xproof}
}

\begin{theorem} 
\label{th:rnd_best_truth}
\label{th:FBRP_r_t}
$f_{\hatv s}^{PR}$ is FBRP from the truth.
\end{theorem}

\jr{
\begin{proof}
We denote the sets of winners and runnerups at time $t$ as $W^t = f^{RP}(\vec a^t); R^t = \{c: s^t(c) = sw^t-1\}$.
We will show by induction that at any step $\vec a^{t-1} \step i \vec a^t$:
\begin{enumerate}
\item $W^t \cup R^t \subseteq W^{t-1} \cup R^{t-1}$ (i.e., candidates not in $W^t \cup R^t$ will not be selected by any agent at a later time).
\item $a_i^t$ is the most preferred candidate for $i$ in $W^{t} \cup R^{t}$ (in particular, a best reply is a direct reply).	
	\item $a^{t-1}_i \succ_i a_i^t$ (in the terminology of \cite{MLR14}, this is a \emph{compromise step}).
\end{enumerate}
Since each voter can make at most $m-1$ compromise steps, convergence is guaranteed within $nm$ steps. 

Assume that for some $t\geq 1$, all of the above holds for any $t'<t$ (so we prove the base case together with the other cases).
Since $\vec a^0$ is truthful, the first step of any voter is always a compromise move. If $i$ had already moved at some previous time $t'<t$, then $a_i^{t'}$ is most preferred in $W^{t'} \cup R^{t'}$. 
	
	By induction, $a=a^{t-1}_i$ is the most preferred candidate in  some $C'$ that contains $W^{t-1} \cup R^{t-1}$ ($C'=C$ in $i$'s first step, and $C'=W^{t'} \cup R^{t'}$ at any other step). Let $x$ and $y$ be $i$'s most preferred candidates in $W^{t-1}$ and in $R^{t-1}$, respectively, and denote the best reply by $a'=a^t_i$. Each of $a$ or $a'$ may belong to $W^{t-1}$, to $R^{t-1}$, or to neither set. This means there are 3X3=9 cases to check. Fortunately, we can show that some of this cases immediately lead to a contradiction, and in the other cases all invariants 1-3 will hold after step $t$.
		
Consider first the case $a\in W^{t-1}$. Since $a$ is most preferred in $C'$, it is strictly more preferred than any other candidate in $W^{t-1}$ or in $R^{t-1}$ (i.e., $a=x$). Thus if $a'\in W^{t-1}$ we get $W^t = \{a'\} \prec_i W^{t-1}$ by Axiom~G. If $a'\in R^{t-1}$ we get $W^t = (W^{t-1}\setminus \{a\})\cup\{a'\} \prec_i W^{t-1}$ by Axiom R. In either case this is not an improvement step for voter~$i$.
	
	Next, suppose $a \notin W^{t-1}$. We further  split to subcases based on $a'$.
	\begin{itemize}
		\item If $a'\in W^{t-1}$ then $f(\vec a_{-i},a')=\{a'\}$. Then $a'=x$, as otherwise  $f(\vec a_{-i},x)=\{x\} \succ_i \{a'\}$, and $i$ is strictly better off by voting for $x$.   This entails $W^t=\{x\},R^{t} = W^{t-1}\setminus \{x\}$ so all invariants 1-3 hold: (1) $W^{t-1} = W^t \cup R^t$; (2) follows from (1) since $a'=x$ is the most preferred in $W^{t-1}$; and (3) follows from (1) since $a=a^{t-1}_i$ is the most preferred in $C'$, and $a'\in C'$. 
		\item If $a' \in R^{t-1}$ then $f(\vec a_{-i},a')=\{a'\} \cup W^{t-1}$. Then $a'=y$, as otherwise $f(\vec a_{-i},y)=\{y\} \cup W^{t-1} \succ_i \{a'\} \cup W^{t-1}$ by Axiom~R, which means $i$ is strictly better off by voting for $y$. This entails $W^t=\{y\} \cup W^{t-1}$, $R^t = R^{t-1}\setminus\{y\}$.  We also get that $a'=y\succ_i x$ or else $x$ would have been a strictly better reply. Thus all invariants 1-3 hold: (1) $W^t = W^{t-1}\cup\{y\}\subseteq W^{t-1}\cup R^{t-1}$ and $R^{t}=R^t\setminus \{y\}$; (2) follows from (1) since $a'=a^t_i=y$ is most preferred in $R^{t-1}$  and strictly preferred to $x$; (3) follows from (1) as in the previous case. 
		\item If $a'\notin  W^{t-1}\cup  R^{t-1}$, then  $W^t=f(\vec a_{-i},a')=W^{t-1}$. The outcome does not change so this cannot be an improvement step for $i$. 
	\end{itemize}
\end{proof}
}
\jr{
\rmr{remove this part? it is the only result where using $\hatv s$ is necessary.
\paragraph{Weighted agents under random tie-breaking}
In contrast to the lexicographic case, the weighted randomized case does not always converge to equilibrium, even with (only) two strategic agents. Moreover, a pure strategy Nash equilibrium may not exist at all.
\begin{proposition} 
\label{ex:rnd_w}
\label{ex:noNE_w_r}
$f^{PR}_{\vec w,\hatv{s}}$ may not have a pure NE (thus does not have any of the finite improvement properties).
\end{proposition}
\begin{xproof}{\ref{ex:rnd_w}}
$C=\{a,b,c\}$, $\hatv{s} =(0,1,3)$. There are 2 agents with weights $w_1=5$, $w_2=3$ and preferences $a\ord 1 b\ord 1 c$, $b \ord 2 c \ord 2 a$ (in particular, $\{b,c\} \ord 1 c$ and  $b \ord 2 \{b,c\} \ord 2 c$). The resulting $3\times 3$ normal form game contains no NE states: (a double arrow means a step to the entry two slots away). 
$$\begin{array}{lccccc}
\text{votes:} & (a,a) & \rightarrow\step{2} & (a,b) &\step{2} & (a,c) \\
\text{scores:} & (8,1,3) \{a\} &          & (5,5,3) \{a\}&         &   (5,1,7)\{c\}\\     
&  \ustep{1} & & \ustep{1} & &\dstep{1}\\ 
        &    (b,a)   &   & (b,b)  &   \lstep{2} & (b,c) \\
				&     (3,6,3) \{b\} &          & (0,9,3) \{b\} &         & (0,6,6) \{b,c\} \\
				&  \uparrow \ustep{1} & & \uparrow \ustep{1} & &\ustep{1}\\ 
				&    (c,a)   & \lstep{2} & (c,b)  &   \lstep{1} & (c,c) \\
				&     (3,1,8) \{c\} &   & (0,4,8) \{c\} &         & (0,1,11) \{c\} \\
\end{array}
$$
\end{xproof} 

}
}
\paragraph{Cardinal utilities}
A (cardinal) utility function is a mapping of candidates to real numbers $u:C\rightarrow \mathbb R$, where $u_i(c)\in \mathbb R$ is the utility of candidate $c$ to agent $i$. We say that $u$ is {\it consistent} with a preference relation $Q_i$ if $u(c)>u(c') \Leftrightarrow c \ord i c'$. 
  The definition of cardinal utility naturally extends to multiple winners by setting $u_i(W)=\frac{1}{|W|}\sum_{c\in W}u_i(c)$ for any subset $W\subseteq C$.\footnote{One interpretation is that we randomize the final winner from the set $W$, and hence the term randomized tie-breaking. For a thorough discussion of cardinal and ordinal utilities in normal form games, see~\cite{Borgers93}.} 

\begin{lemma} 
\label{lemma:rel} Consider any cardinal utility function $u$ and the partial preference order $Q$ it induces on subsets by random tie-breaking.  
$Q$ holds Axioms~K+G+R. 
\end{lemma}
The proof is rather straight-forward, and is deferred to the appendix. 
%

\jr{
\rmr{maybe omit}
\begin{proposition}
\label{ex:rnd_better}
\label{ex:no_FIP_r_t}
$f^{PR}$  is not FIP from the truth.
\end{proposition}

\begin{xproof}{~\ref{ex:rnd_better}}
We use 5 candidates with initial score $\hatv s=(1,1,2,0,0)$, and 2 agents with utilities $u_1 = (5,3,2,8,0)$ and $u_2 = (4,2,5,0,8)$. In particular, $\{b,c\}\ord 1 c$, $ \{a,c\}\ord 1 \{a,b,c\}$, and $\{a,b,c\} \ord 2 \{b,c\}$, $ c\ord 2 \{a,c\}$, 
 and the following cycle occurs:
$(1,1,2,1,1) \{c\} \step{1} (1,2,2,0,1) \{b,c\} \step{2} (2,2,2,0,0) \{a,b,c\} \step{1} (2,1,2,1,0) \{a,c\} \step{2} (1,1,2,1,1) \{c\}$.
\end{xproof}
}

Finally, in contrast to the lexicographic case, convergence is no longer guaranteed if agents start from an arbitrary profile of votes, or are allowed to use direct-replies that are not best-replies.
\jr{The following example shows that in the randomized tie-breaking setting even direct best reply dynamics may have cycles, albeit for specific utility scales.
}
\begin{proposition} 
\label{ex:no_RFIP_r}
$f^{PR}$ is not restricted-FIP.
\end{proposition}

\begin{xproof}{~\ref{ex:no_RFIP_r}}
There are 4 candidates $\{a,b,c,x\}$ and 3 agents with utilities $u_1 = (7,3,0,4)$, $u_2 = (0,7,3,4)$ and $u_3 = (3,0,7,4)$. In particular, the following preference relations hold: $a \ord 1 \{a,b\} \ord 1 x \ord 1  \{a,c\}$; $b \ord 2 \{b,c\} \ord 2 x \ord 2  \{a,b\}$; and $c \ord 3 \{a,c\} \ord 3 x \ord 3  \{b,c\}$. 

Consider the initial state $\vec a_0 = (a,b,x)$ with $\vec s(\vec a_0) = (1,1,0,1)$ and the outcome $\{a,b,x\}$. We have the following cycle where every step is the unique  reply of the playing agent.
\jr{
$$
\begin{array}{ccccc}
(1,1,0,1) \{a,b,x\} &\step{2} &(1,0,0,2) \{x\} &\step{3} &(1,0,1,1) \{a,x,c\} \\
\ustep{1} & & & & \dstep{1} \\
(0,1,0,2) \{x\}  & \lstep{3} &(0,1,1,1) \{x,b,c\} & \lstep{2} & (0,0,1,2) \{x\} 
\end{array}
$$
}
\end{xproof}

\begin{proposition}{}
\label{ex:no_FDRP_r_t}
 $f^{PR}$ is not FDRP even from the truth.
\end{proposition}
\begin{xproof}{~\ref{ex:no_FDRP_r_t}}
We take the game from Ex.~\ref{ex:no_RFIP_r}, and add for each voter $i\in \{1,2,3\}$ a candidate $d_i$, s.t. $u_i(d_i)=8, u_i(d_j)=j$ for $j\neq i$. We also add an initial score of $3$ to each of the candidates $\{a,b,c,x\}$. 
Voter~3 moves first to $a_3^1=x$, which is a direct reply. Then voters 1 and 2 move to their best replies $a,b$, respectively. Now the cycle continues as in  Ex.~\ref{ex:no_RFIP_r}. 
\end{xproof}

\subsection{Stochastic Dominance and Local Dominance}\label{subsec:SD}
While assigning cardinal utilities is one way to deal with ties, it is sometimes preferable not to assume a particular cardinal utility scale. Denote by $f^P(\vec a)\subseteq C$ the subset of candidates with maximal Plurality score, before any tie-breaking takes place.  We can still derive a well-defined dynamics from any partial order over subsets of candidates, by assuming that a voter performs a better-response step if she strictly prefer the new outcome, and otherwise (if the new outcome is same, worse, or incomparable) she does not move.

One example of such a partial order is \emph{stochastic dominance}  (SD), which was applied to tie-breaking by \cite{reyhani2012best}. A different partial order is implied by \emph{local dominance} (LD) which was defined for voting with uncertainty about the outcome~\cite{conitzer2011dominating,MLR14}, when uncertainty is regarding the tie breaking. We show how convergence results for LD/SD dynamics fit with other results. 
%
\paragraph{Stochastic dominance}  Reyhani and Wilson assume that ties are broken uniformly at random, and that a voter will only perform a step that stochastically dominates the current winner(s), if such exists.

\begin{theorem}[Theorem~5.7 in \cite{reyhani2012best}]\label{th:SD}
Plurality with stochastic dominance tie-breaking  is FDBRP.
\end{theorem}

We can show the following (see appendix):
\begin{proposition}
\label{th:SD_axioms}
 A step $\vec a\step i \vec a'$ is a better-response under random tie-breaking and stochastic dominance, if and only if $f^P(\vec a') \succ_i f^P(\vec a)$ is entailed by $Q_i$,  Axioms~K+G+R, and transitivity.
\end{proposition}
In other words, while Theorem~\ref{th:FBRP_r_t} allowed any moves \emph{consistent} with the axioms, SD allows only moves that \emph{follow} from the axioms,  and explicitly forbid any other step.  Thus it is more restricted than expected-utility based randomized tie-breaking.

Since any SD step is also a better-reply under any cardinal utility scale, any strong or restricted convergence result for the latter applies to the former, but not vice-versa.

\paragraph{Local dominance}
Suppose that there are several candidates with maximal score. 
A voter may consider all of them as ``perhaps winners,'' without specifying how the actual winner is selected.
If the voter is concerned about making a move that will leave her worse off, she will only make moves that will improve her utility with certainty, i.e. that dominates her current action (where possible worlds are all strict tie-breaking orders)~\cite{conitzer2011dominating,MLR14,Meir15}.\footnote{\cite{MLR14,Meir15} consider more general uncertainty over candidates' score, and \cite{conitzer2011dominating} considers arbitrary information sets.}

\begin{theorem}[Theorem~11 in the full version of \cite{Meir15}]
Plurality with Local-Dominance tie-breaking is FDRP.
\end{theorem}

To see how this compares with other convergence results, we need the following proposition (see appendix). 

\begin{proposition}
\label{th:LD_axioms}
 A step $\vec a\step i \vec a'$ is a better-response under unknown tie-breaking and local dominance, if and only if $f^P(\vec a') \succ_i f^P(\vec a)$ is entailed by $Q_i$,  Axioms~K+G, and transitivity.
\end{proposition}

Note that since Axioms~K+G+R include K+G, any LD step is also an SD step, so a restriction to LD can only eliminate cycles. Thus FBDRP follows from Theorem~\ref{th:SD}.
We note that with either SD or LD tie-breaking there may be new stable states that are not Nash-equilibria. Even so, an analysis of Ex.~\ref{ex:noFIP_r} shows that all steps are entailed by Axioms~K+G (and thus by Axioms~K+G+R). Thus neither game form is FIP. 

\medskip
What if we assume that voters are even more risk-averse and only follow steps that are better-replies by Axiom~K? Then it is easy to see that only moves to a more-preferred candidate can be better-replies (any move to or from a tie cannot follow from Axiom~K and is thus forbidden), which means there are trivially no cycles.

\ssection{Weak Acyclicity}\label{sec:weak}
Except for Plurality and Veto, convergence is not guaranteed even under restrictions on the action scheduler and the initial state. In contrast, simulations~\cite{grandi2013restricted,MLR14,koolyk2016convergence} show that iterative voting almost always converges even when this is not guaranteed by theory. 
We believe that weak acyclicity is an important part of the explanation to this gap.

\subsection{Plurality  with Random tie-breaking}

We have seen in Section~\ref{sec:Plurality} that while $f^{PR}$ is FDRP from the truthful initial state, this is no longer true from arbitrary states, and in fact $f^{PR}$ is not restricted-FIP under any action scheduler. 
Our main theorem in this section shows that under a certain scheduler (of agents+actions), convergence is guaranteed from \emph{any} state. Further, this still holds if actions are restricted to direct-replies. 

\begin{nlemma}
\label{lemma:large_gap}
Consider any game $G=\tup{f^{PR}_{\hatv s},\vec Q}$.
Consider some candidate $a^*$, and suppose that in $\vec a^0$, there are $x,y$ s.t. $s^0(x)\geq s^0(y) \geq s^0(a^*)+2$.
	Then for any sequence of direct replies, $a^*\notin f(\vec a^t)$. 
	\end{nlemma}
	\begin{proofnoqed}
	We show that at any time $t\geq 0$ there are $x^t,y^t$ s.t.  $s^0(x),s^0(y) \geq s^0(a^*)+2$.
	For $t=0$ this holds for $x^t=x,y^t=y$. 
	Assume by induction that the premise holds for $\vec a^{t-1}$. Then there are two cases:
	\begin{enumerate}
		\item $|f(\vec a^{t-1})|\geq 2$. Then since step $t$ must be a direct reply, it must be to some candidate $z$ with $s^{t-1}(z)\geq sw^{t-1}-1$. Also, either $x^{t-1}$ or $y^{t-1}$ did not lose votes (w.l.o.g. $x^{t-1}$). Thus $s^t(x),s^t(z)\geq sw^{t-1} \geq s^{t-1}(a^*)+2 \geq s^{t}(a^*)+2$.
		\item $|f(\vec a^{t-1})|= 1$. Then suppose $f(\vec a^{t-1})=\{x^{t-1}\}$, and we have that $sw^{t-1}\geq s^{t-1}(a^*)+3$. The next step is $z$ where either $s^{t-1}(z)=sw^{t-1}-1$ (and then we conclude as in case~1), or $s^{t-1}(z)=sw^{t-1}-2$ and $x^{t-1}$ loses 1 vote. In the latter case, $s^t(x^{t-1})=s^t(z)=sw^{t-1}-1 \geq s^{t-1}(a^*)+2 \geq s^{t}(a^*)+2$.
		\qed
	\end{enumerate}
	
	\end{proofnoqed}

\begin{theorem}
\label{th:weakFDRP_r}
$f^{PR}_{\hatv s}$ is weak-FDRP.
\end{theorem}

\begin{proof}
Consider a game $G=\tup{f^{PR}_{\hatv s},\vec Q}$, and an initial state $\vec a^0$. 
For a state $\vec a$, denote by $B(\vec a)  \subseteq A^n$ all states reachable from $\vec a$ via paths of direct replies.
Let $B=B(\vec a^0)$, and assume towards a contradiction that $B$ does not contain a Nash equilibrium. For every $\vec b\in B$, let $C(\vec b)= \{c\in C: \exists \vec a\in B(\vec b) \wedge c\in f(\vec a)\}$, i.e. all candidates that are winners in some state reachable from $\vec b$. 

For any $\vec b\in B(\vec a^0)$, define a game $G_{\vec b}$ by taking $G$ and eliminating all candidates \emph{not in} $C(\vec b)$. Since we only consider direct replies, for any $\vec a\in B(\vec b)$, the set of outgoing edges $I(\vec a)$ is the same in $G$ and in $G_{\vec b}$ (as any direct reply must be to candidate in $C(\vec b)$). Thus by our assumption, the set $B(\vec b)$ in game $G_{\vec b}$ does not contain an NE.

For any $\vec b\in B(\vec a^0)$, let $\vec b^*$ be the truthful state of game $G_{\vec b}$, and let $T(\vec b)\subseteq N$ be the set of agents who are truthful in $\vec b$. That is, $i\in  T(\vec b)$ if $b_i = b^*_{i}$. 

Let $\vec b^0$ be some state  $\vec b\in B(\vec a^0)$ s.t. $|T(\vec b)|$ is maximal, and let $T^0=T(\vec b^0)$. If $|T^0|=n$ then $\vec b^0$ is the truthful state of $G_{\vec b^0}$, and thus by Theorem~\ref{th:FBRP_r_t} all  best-reply paths from $\vec b^0$ in $G_{\vec b^0}$ lead to an NE, in contradiction to $B(\vec b^0)$ not containing any NE. Thus  $T^0<n$.  We will prove that there is a path from $\vec b^0$ to a state $\vec b'$ s.t. $|T(\vec b')|>|T^0|$. 

Let $i\notin T(\vec b^0)$ (must exist by the previous paragraph). Consider the score of candidate $b^*_i$ at state $\vec b^0$. We divide into 5 cases. All scores specified below are in the game $G_{\vec b^0}$.

 \begin{enumerate}[label*=Case~\arabic*.]
	 \item $|f(\vec b^0)|>1$ and $b^*_i\in f(\vec b^0)$ (i.e $b^*_i$ is one of several winners). Then consider the step $\vec b^0 \step i b^*_i$. This make $b^*_i$ the unique winner, and thus it is a direct best-reply for $i$. In the new state $\vec b'=(\vec b^0_{-i},b^*_i)$ we have $T(\vec b') = T(\vec b^0) \cup \{i\}$. 
	\item $s^0(b^*_i) = sw^0-1$ (i.e., $b^*_i$ needs one more vote to become a winner). By Axioms~G+R, $i$ prefers $f(\vec b^0_{-i},b^*_i)$ over $f(\vec b^0)$. Then similarly to case 1, $i$ has a direct step $\vec b^0 \step i b^*_i$, which results in a ``more truthful'' state $\vec b'$. \rmr{this may not be a best-reply}
	\item $b^*_i =  f(\vec b^0)$  (i.e $b^*_i$ is the unique winner). Then the next step $\vec b^0 \step j \vec b^1$ will bring us to one of the two previous cases. Moreover, it must hold that $j\notin T(\vec b^0)$ since otherwise $b^0_j = b^*_j = f(\vec b^0)$ which means $I_j(\vec b^0)=\emptyset$. Thus $|T(\vec b')| = |T(\vec b^1)|+1 \geq |T(\vec b^0)|+1$.
	\item $f(\vec b^0)=x\neq b^*_i$, and $s^0(x) = s^0(b^*_i)+2$. We further divide into:
	\begin{enumerate}[label*=\arabic*.]
		\item $s^0(b^*_i)\geq s^0(y)$ for all $y\neq x$. Then the next step by $j$ must be from $x$, which brings us to one of the two first cases (as in Case~3).
		\item There is $y\neq x$ s.t. $s^0(x)=s^0(y)+1=s^0(b^*_i)+2$. Then we continue the sequence of steps until the winner's score decreases. Since all steps that maintain $sw^t$ select a more preferred candidate, this most occur at some time $t$, and $T(\vec b^0)\subseteq T(\vec b^t)$. Then at $\vec b^t$ we are again in Case~1 or 2.
		\item There is $y\neq x$ s.t. $s^0(x)=s^0(y)=s^0(b^*_i)+2$. Then by Lemma~\ref{lemma:large_gap} $b^*_i$ can never be selected, in contradiction to $b^*_i\in C(\vec b^0)$. 
	\end{enumerate}
	\item $f(\vec b^0)=x\neq b^*_i$, and $s^0(x) \geq s^0(b^*_i)+3$. We further divide into:
	\begin{enumerate}[label*=\arabic*.]
			\item For all $y\neq x$, $s^0(y)\leq s^0(x)-3$. In this case no reply is possible.
			\item There is some $y\neq x$ s.t. $s^0(y)\geq s^0(b^*_i)+2$. Then  by Lemma~\ref{lemma:large_gap} $b^*_i$ can never be selected, in contradiction to $b^*_i\in C(\vec b^0)$. 
			\item There is some $y\neq x$ s.t. $s^0(y)\geq s^0(b^*_i)+1$ Then the next step must be from $x$ to such $y$. Which means $s^1(x)=s^1(y)=sw^0-1 \geq s^0(b^*_i)+2 = s^1(b^*_i)+2$. Thus again by Lemma~\ref{lemma:large_gap} we reach a contradiction. 
	\end{enumerate}	
 \end{enumerate}
Therefore we either construct a path of direct replies to $\vec b'\in B(\vec b^0)$ with $|T(\vec b')|>|T(\vec b^0)|$ in contradiction to our maximality assumption, or we reach another contradiction. Thus $B(\vec b^0)$ must contain some NE (both in $G_{\vec b^0}$ and in $G$), which means by construction that $G$ is weakly-FDRP from $\vec b^0$. However since $\vec b^0\in B(\vec a^0)$, we get that $G$ is weakly-FDRP from $\vec a^0$ as well. \qed
\end{proof}
\jr{
\begin{remark}\label{rem:Kuk1}
Theorem~\ref{th:weakFDRP_r} and Ex.~\ref{ex:no_RFIP_r} provide  a partial answer to an open question regarding whether there are game forms that admit weak FIP but not restricted FIP~\cite{kukushkin2011acyclicity}. Indeed, the game form $f^{PR}$ for $m=4,n=3$ is such an example, but one that uses randomization. However if we think of $f^{PR}$  as a deterministic game form with $2^m-1$ possible outcomes (all nonempty subsets of candidates), where players are restricted to $m$ actions each, then the allowed utility profiles are constrained (by Axioms~G and R) and thus this result does not settle Kukushkin's question completely. 
  \end{remark}
}

\subsection{Weighted Plurality}
\label{sec:weak_weighted}
When voters are weighted, cycles of direct responses can emerge~\cite{MPRJ:2010:AAAI,Meir:2016:COMSOC}. We conjecture that such cycles must depend on the order of agents, and that certain orders will break such cycles and reach an equilibrium, at least from the truthful state.
\begin{conjecture}
$f^{PL}_{\hatv s,\vec w}$ is weak-FDRP (in particular weak-FIP).
\end{conjecture}
Similar techniques to those used so far appear to be insufficient to prove the conjecture. For example, in contrast to the unweighted case, a voter might return to a candidate she deserted in \emph{any scheduler}, even if only two weight levels are present. 
We thus leave the proof of the general conjecture for future work.

Yet, we want to demonstrate the power of weak acyclicity over restricted acyclicity, even when there are  no randomness or restrictions on the utility space. That is, to provide a definite (negative) answer to Kukushkin's question of whether weak acyclicity entails restricted acyclicity. To do so, we will use a slight variation of Plurality with weighted voters and lexicographic tie-breaking.

\begin{figure}[t]
\input{f_star.tex}
\caption{\label{fig:f_star}In each state we specify the actions of all 3 agents, and the outcome in curly brackets. Agent~1 controls the horizontal axis, agent~2 the vertical axis, and agent~3 the in/out axis. We omit edges between states with identical outcomes, since such moves are impossible for any transitive preferences.
 A directed edge in (b) is a better-reply in $G^*$. }
\end{figure}

\begin{theorem}
\label{th:f_star}
There exist a game form $f^*$ s.t. $f^*$ is weak-FIP but not restricted-FIP.
\end{theorem}
\begin{proof}
Consider the following game $G$:
The initial fixed score of candidates $\{a,b,c,d\}$ is $\hatv s = (0,1,2,3)$. The weight of each voter $i\in \{1,2,3\}$ is $i$. The preference profile is as follows: $c \ord 1 d \ord 1 b \ord 1 a$, $ b \ord 2 c \ord 2 a \ord 2 d$,  and $a \ord 3 b \ord 3 c \ord 3 d$.
This game was used in \cite{MPRJ:2010:AAAI} to demonstrate that Plurality with weighted voters is not FDRP, however it can be verified that $G$ is restricted-FIP so it is not good enough for our use.

If we ignore agents' preferences, we get a particular game form $f^{PL}_{\hatv s,\vec w}$ where $N=\{1,2,3\}$, $M=\{a,b,c,d\}$, $\hatv s = (0,1,2,3)$ and $\vec w=(1,2,3)$. 

We define $f^*$ by modifying $f^{PL}_{\hatv s,\vec w}$ with the following restrictions on agents' actions: $A_1=\{c,d\}, A_2=\{b,c\}, A_3=\{a,b,d\}$. Thus $f^*$ is a $2 \times 2 \times 3$ game form, presented in Figure~\ref{fig:f_star}(a).

We first show that $f^*$ is not restricted-FIP. Indeed,  consider the game $G^*$ accepted from $f^*$ with the same preferences from game $G$ (Figure~\ref{fig:f_star}(b)). We can see that there is a cycle of length 6 (in bold). An agent scheduler that always selects the agent with the bold reply guarantees that convergence does not occur, since in all 6 relevant states the selected agent has no alternative replies.

Next, we show that $f^*$ is weak-FIP. That is, for any preference profile there is some scheduler that guarantees convergence. We thus divide into cases according to the preferences of agent~3. 
In each case, we specify a state where the scheduler selects agent~3, the action of the agent, and the new state. 

We note that since all thick edges must be oriented in the same direction, $a \succ_3 b$ if and only if $b \succ_3 c$. Thus the following three cases are exhaustive. 

\begin{tabular}{c|c|c|c|c}
& $Q_3$ & state & action & new state \\
\hline
1 &$b \succ d$ & $(d,b,a)$ & $b$ & $(d,b,b)$ \\
\hline
2 & $d \succ b ~\&~ d\succ a$ & $(c,b,b)$ & $d$ & $(c,b,d)$ \\
\hline
3& $a\succ d \succ b\succ c$ & $(d,c,b)$ & $d$ & $(d,c,d)$ \\
\end{tabular}

In either case, agent~3 moves from a state on the cycle to a Nash equilibrium. 
\end{proof}

\ssection{Conclusions and Future Work}
\label{sec:conclusion}

The main conceptual contribution of this work was to provide a joint rigorous framework for the study of iterative voting, as part of the broader literature on acyclicity of games and game forms. 

On the technical level, this unified presentation enabled us to construct examples of voting rules that settle at two open questions on acyclicity of game forms: first, showing that there may be non-separable game forms that are FIP (Theorem~\ref{th:FIP}); and second, that there are game forms that are weakly acyclic but not order-free acyclic (Theorem~\ref{th:f_star}). 

In addition, we provide an extensive study of convergence properties of the common Plurality rule and its variations. 
We summarize all known results on iterative voting that we are aware of in Table~\ref{tab:results_rules}. Note that in some cases we get positive results if we restrict the initial state or the number of voters (not shown in the table). For Plurality we provide a more detailed picture in Figs.~\ref{fig:FIP},\ref{fig:FIP_r}. 
Previous papers whose results are covered in the Table~\ref{tab:results_rules} often use different terminology and thus theorems and examples need to be rephrased (and sometimes slightly modified) to be directly comparable. These rephrasing and necessary modifications are explained in detail in \cite{Meir:2016:COMSOC}. The only paper not covered in \cite{Meir:2016:COMSOC} is by Koolyk et al.~\shortcite{koolyk2016convergence}, which provided non-convergence examples for a variety of common voting rules including \emph{Maximin, Copeland, Bucklin, STV, Second-Order Copeland, and Ranked Pairs}. All results demonstrate cycles under best-reply (and under several other restrictions) from the truthful state, thereby proving that neither of these rules is FBRP (even from the truth).

\input{result_tables}

\begin{figure}[t]		
		\small{
					$$\begin{tabular}{c|ccc|ccccc}
  FBRP (Ex.~\ref{ex:no_FBRP}) \XX&  && & restricted-FBRP     \VV         & $\Rightarrow$ &  weak-FBRP \VV &&\\
	      $\Uparrow$ & &     &&      $\Downarrow$      & &   $\Downarrow$      &&\\
	       FIP \XX & &        &&   restricted-FIP   \VV & $\Rightarrow$ & weak-FIP \VV &\\		
					              $\Downarrow$      & & && $\Uparrow$ & &   $\Uparrow$ & &  \\
				FDRP  (Thm.~\ref{th:FDRP}) \VV & $\Rrightarrow$ & FDBRP \VV & $\Rrightarrow$ & restricted-FDRP   \VV            & $\Rightarrow$ &  weak-FDRP \VV &&\\
				 \end{tabular}$$}
				\caption{\label{fig:FIP} Convergence results for Plurality under lexicographic tie-breaking. Positive results (in light green) carry with the direction of the arrows, whereas negative results (dark gray) carry in the opposite direction. 
				}
				\end{figure}

								\begin{figure}[t]
								\small{
					$$\begin{tabular}{ccc|c|ccccc}
FBRP from truth (Thm.~\ref{th:FBRP_r_t}) \VV &$\Leftarrow$& FBRP \XX  & & restricted-FBRP         \XX      & $\Rightarrow$ &  weak-FBRP ? &&\\
	$\Uparrow$ &&      $\Uparrow$ & &           $\Downarrow$      & &   $\Downarrow$      &&\\
	      FIP from truth \XX  &$\Leftarrow$& FIP \XX & $\Rrightarrow$&           restricted-FIP  (Ex.~\ref{ex:no_RFIP_r})  \XX & $\Rightarrow$ & weak-FIP \VV & &\\	
					 $\Downarrow$  &&           $\Downarrow$      & &  $\Uparrow$ & &   $\Uparrow$ & &  \\
			FDRP from truth (Ex.~\ref{ex:no_FDRP_r_t})\XX	 &$\Leftarrow$& FDRP \XX & & restricted-FDRP           \XX    & $\Rightarrow$ &  weak-FDRP (Thm.~\ref{th:weakFDRP_r})\VV &&\\
			   \end{tabular}$$
				}
				%

\caption{\label{fig:FIP_r} Convergence results for Plurality under random tie-breaking.\vspace{-2mm}}
				
\end{figure}

Beyond the direct implication of various acyclicity properties on convergence in an interactive setting where agents vote one-by-one, [strong/weak] acyclicity is tightly linked to the convergence properties of more sophisticated learning strategies in repeated games~\cite{bowling2005convergence,marden2007regret}, which is another reason to understand them.

Fabrikant et al.~\shortcite{fabrikant2010structure} provide a sufficient condition for weak-acyclicity, namely that any subgame contains a \emph{unique} Nash equilibrium. Unfortunately, this criterion is not very useful for most voting rules, where typically (at least) all unanimous votes form equilibria. 
Another sufficient condition due to Apt and Simon~\shortcite{apt2012classification} is by eliminating never-best-reply strategies, and the prospects of applying it to common voting rules is not yet clear.

\jr{
 We can see that in the ``standard'' lexicographic domain, convergence is guaranteed from any initial state provided that voters restrict themselves to direct replies. With randomized tie-breaking, we must also require a truthful initial vote. On the other hand, we can also allow indirect best-replies, so the results are essentially  incomparable. However, we see the result in the lexicographic case as stronger, since it only requires a very mild and natural behavioral restriction in the context of Plurality voting, whereas it is harder to justify assumptions on the initial state.
}

\paragraph{Implications on social choice}
Importantly, best-reply dynamics is a natural and straightforward process, and requires little information. As such, and due to the convergence properties demonstrated in this work, it is an attractive ``baseline'' candidate for predicting human voter behavior in elections and designing artificial agents with strategic voting capabilities---two of the most important, and also the hardest,  goals of social choice research. However, the clear disadvantage of this approach is that in the vast majority of cases (especially when there are more than a handful of voters), almost every voting profile (including the truthful one) is already a Nash equilibrium. 
Given this, our analysis is particularly suitable when the number of voters is small, for two main reasons. First, it is more practical to perform an iterative voting procedure with few participants. Second, the question of convergence is only relevant when cases of tie or near-tie are common.  
In more complex situations with many active voters who may change their vote, it is likely that a more elaborate game-theoretic model is required, which takes into account voters' uncertainty and heuristic behavior (see Section~\ref{sec:related}). 

%

\paragraph{Promising future directions}

Based on the progress made in this paper and the other results published since the introduction of iterative voting in \cite{MPRJ:2010:AAAI}, 
 we believe that research in this area should focus on three primary directions:
\begin{enumerate}
	\item Weak-acyclicity seems more indicative than order-free acyclicity to determine convergence in practice. Thus theorists should study which voting rules are weak-FIP, perhaps under reasonable restrictions (as we demonstrated, this property is distinct from restricted-FIP). We highlight that even in rules where there are counter examples for weak acyclicity (k-approval, Borda), these examples use two voters and games with more voters may well be weakly acyclic. 
\item It is important to experimentally study how people really vote in iterative settings (both in and out of the lab), so that this behavior can be formalized and behavioral models can be improved. The work of \cite{TMG15} is a preliminary step in this direction, but there is much more to learn. Ideally, we would like to identify a few types of voters, such that for each type we can relatively accurately predict the next action in a particular state. It would be even better if these types are not specific to a particular voting rule or contextual details. 
\item We would like to know not only if a voting rule converges under a particular dynamics (always or often), but also what are the properties of the attained outcome---in particular, whether the iterative process improves welfare or fairness, avoids ``voting paradoxes''~\cite{xia2007sequential} and so on. Towards this end, several researchers (e.g.,~\cite{RE12,branzeibad,MLR14,bowman2014potential,koolyk2016convergence}) have started to explore these questions via theory and simulations. 
However, a good understanding of how iterative voting shapes the outcome, whether the population of voters consists of humans or artificial agents, is still under way. 
\end{enumerate}

%


\bibliographystyle{named}
\bibliography{abbshort,plurality.aaai,ultimate} 
\newpage
\appendix
\input{plurality.TEAC.appendix}

\end{document}

%% file: defs.tex
\floatname{algorithm}{Mechanism}
\newtheorem{theorem}{Theorem}

\newtheorem{lemma}[theorem]{Lemma}

\newtheorem{proposition}[theorem]{Proposition}

\newtheorem{conjecture}[theorem]{Conjecture}

\newtheorem{remark}{Remark}[section]
\newtheorem{definition}{Definition}[section]

\newenvironment{rproposition}[1]{\medskip\noindent\textbf{Proposition~\ref{#1}.}\begin{itshape}}{\end{itshape}\medskip}
\newenvironment{rlemma}[1]{\medskip\noindent\textbf{Lemma~\ref{#1}.}\begin{itshape}}{\end{itshape}\medskip}

\newcommand{\sproof}{\noindent{\bf Proof.}\hspace*{1em}}

\newenvironment{proofnoqed}{\sproof}{}

\def\literalqed{{\ \nolinebreak\hfill\mbox{\quad}}}

\long\def\symbolfootnote[#1]#2{\begingroup%
\def\thefootnote{\fnsymbol{footnote}}\footnote[#1]{#2}\endgroup} 

    \makeatletter
    \renewcommand\part{%
      \if@openright
        \cleardoublepage
      \else
        \clearpage
      \fi
      \thispagestyle{empty}%
      \if@twocolumn
        \onecolumn
        \@tempswatrue
      \else
        \@tempswafalse
      \fi
      \null\vfil
      \secdef\@part\@spart}
    \makeatother

\def\cite{\citep}

\def\shortcite{\citeyearpar}

\def\calA{{\cal A}}

\def\calB{{\cal B}}

\DeclareMathOperator{\argmax}{\text{argmax}}

\newcommand{\ord}[1]{\succ_{#1}}

\newcommand{\labeq}[2]{\begin{equation}\label{eq:#1}#2\end{equation}}

\newcommand{\ceil}[1]{\left\lceil #1 \right\rceil}

\def\({\left(}
\def\){\right)}

\newcommand{\tup}[1]{\left\langle #1 \right\rangle}

\newcommand{\xqed}{\mbox{\raggedright $\Diamond$}}
\def\xqedhere{\hfill\xqed}
\newenvironment{xproof}[1]{\noindent\textit{Example~#1.}}{\xqedhere\vspace{1ex}}

\newcommand{\step}[1]{\stackrel{{\scriptscriptstyle{#1}}}{\rightarrow}}
\newcommand{\lstep}[1]{\stackrel{{\scriptscriptstyle{#1}}}{\leftarrow}}
\newcommand{\ustep}[1]{\uparrow{\scriptscriptstyle{#1}}}
\newcommand{\dstep}[1]{\downarrow{\scriptscriptstyle{#1}}}

\def\newpar{\vspace{-3mm}\paragraph}

\renewcommand{\vec}{\mathbf}

%% file: example_proof_DFRP.tex
%


\begin{figure}
\centering
\begin{tabular}{rl}
\begin{tikzpicture}[scale=0.9,transform shape]

\tikzstyle{vote1}=[draw=black,fill=white,
inner sep=0pt,minimum size=5mm]
\tikzstyle{votex}=[draw=white,fill=white,
inner sep=0pt,minimum size=5mm]
\tikzstyle{vote2}=[draw=black,fill=black!10!white,
inner sep=0pt,minimum size=5mm]
\tikzstyle{vote3}=[draw=black,fill=black!30!white,
inner sep=0pt,minimum size=5mm]
\tikzstyle{vote4}=[draw=black,fill=black!50!white,
inner sep=0pt,minimum size=5mm]
\tikzstyle{vote5}=[draw=black,fill=black!70!white,text=white,
inner sep=0pt,minimum size=5mm]	

%
	%
	%
	%

	\draw[thick] (1.5,1) -- (6.5,1);	

	\node at (2,0.7) {};
	\node at (3,0.7) {$b$};
	\node at (4,0.7) {$cw^{t-1}$};
	\node at (5,0.7) {};
	\node at (6,0.7) {$a$};
	
	\node at (3,4.5) [votex] {};
	
	\draw (1.75,1) rectangle (2.25,2.5);

	\draw (2.75,1) rectangle (3.25,4);
	
		\draw (3.75,1) rectangle (4.25,4);
		
			\draw (4.75,1) rectangle (5.25,2);	
			
			\draw (5.75,1) rectangle (6.25,3.5);

	    \node at (6,3.75) [vote1] {$i$};
			
	\draw[dashed] (1.5,4) -- (6.5,4);
		\node at (7,4) {$\overline s$};

\end{tikzpicture}
&

\begin{tikzpicture}[scale=0.9,transform shape]

\tikzstyle{vote1}=[draw=black,fill=white,
inner sep=0pt,minimum size=5mm]
\tikzstyle{votex}=[draw=white,fill=white,
inner sep=0pt,minimum size=5mm]
	\draw[thick] (1.5,1) -- (6.5,1);	

	\node at (2,0.7) {};
	\node at (3,0.7) {$b=cw^t$};
	\node at (4,0.7) {};
	\node at (5,0.7) {};
	\node at (6,0.7) {$a$};
	
	\node at (3,4.5) [votex] {};
	
	\draw (1.75,1) rectangle (2.25,2.5);

	\draw (2.75,1) rectangle (3.25,4);
	
		\draw (3.75,1) rectangle (4.25,4);
		
			\draw (4.75,1) rectangle (5.25,2);	
			
			\draw (5.75,1) rectangle (6.25,3.5);

	    \node at (3,4.25) [vote1] {$i$};
			
	\draw[dashed] (1.5,4) -- (6.5,4);
		\node at (7,4) {$\overline s$};
		\node at (6.75,3.5) {$s^t(a)$};
		\node at (2.3,4.4) {$s^t(b)$};
\end{tikzpicture}
\end{tabular}

\caption{\label{fig:proof1}An illustration of a type~1 move. Tie-breaking is in favor of the left most candidate.}
\end{figure}


%% file: f_star.tex
%
\centering
\subfloat[The game form $f^*$]{ 
\begin{tikzpicture}[scale=0.75,transform shape]
  \node [label=center:{$(c\,c\,d)\{d\}$}](ccd) at (0,0) {};
	\node [label=center:{$(c\,c\,b)\{c\}$}](ccb) at (2,1) {};
	\node [label=center:{$(c\,c\,a)\{c\}$}](cca) at (4,2) {};
	\node[label=center:{$(d\,c\,d)\{d\}$}](dcd) at (12,0) {};
	\node[label=center:{$(d\,c\,b)\{b\}$}](dcb) at (10,1) {};
	\node[label=center:{$(d\,c\,a)\{c\}$}](dca) at (8,2) {};
	\node[label=center:{$(c\,b\,d)\{d\}$}](cbd) at (0,6) {};
	\node[label=center:{$(c\,b\,b)\{b\}$}](cbb) at (2,5) {};
	\node[label=center:{$(c\,b\,a)\{a\}$}](cba) at (4,4) {};
	\node[label=center:{$(d\,b\,d)\{d\}$}](dbd) at (12,6) {};
	\node[label=center:{$(d\,b\,b)\{b\}$}](dbb) at (10,5) {};
	\node[label=center:{$(d\,b\,a)\{d\}$}](dba) at (8,4) {};
	
  %
  \tikzstyle{LabelStyle}=[fill=none]
	\tikzstyle{EdgeStyle}=[above,shorten >= 0.25cm, shorten <=0.25cm]
  \Edge[label=$3$](ccd)(ccb)
	
	\Edge[label=$3$](dcd)(dcb)
	\Edge[label=$3$](dcb)(dca)
	
	\Edge[label=$3$](cbd)(cbb)
	\Edge[label=$3$](cbb)(cba)
	
	\Edge[label=$3$](dbd)(dbb)
	\Edge[label=$3$](dbb)(dba)
	
	\tikzstyle{EdgeStyle}=[above,shorten >= 0.5cm, shorten <=0.5cm]
	\Edge[label=$1$](ccb)(dcb)
	\Edge[label=$1$](cba)(dba)
	
	\tikzstyle{EdgeStyle}=[left,shorten >= 0.15cm, shorten <=0.15cm]
	\Edge[label=$2$](cbb)(ccb)
	\Edge[label=$2$](cba)(cca)
	\Edge[label=$2$](dba)(dca)

   \tikzstyle{EdgeStyle}=[bend left, above,shorten >= 0.2cm, shorten <=0.5cm]
	\Edge[label=$3$](cbd)(cba)
	\Edge[label=$3$](dca)(dcd)
\end{tikzpicture}
}
\\
\vspace{-4mm}
\subfloat[The game $G^*$]{\label{sfig:G2}
\begin{tikzpicture}[scale=0.75,transform shape]
  \node [label=center:{$(c\,c\,d)\{d\}$}](ccd) at (0,0) {};
	\node [label=center:{$(c\,c\,b)\{c\}$}](ccb) at (2,1) {};
	\node [label=center:{$(c\,c\,a)\{c\}$}](cca) at (4,2) {};
	\node[label=center:{$(d\,c\,d)\{d\}$}](dcd) at (12,0) {};
	\node[label=center:{$(d\,c\,b)\{b\}$}](dcb) at (10,1) {};
	\node[label=center:{$(d\,c\,a)\{c\}$}](dca) at (8,2) {};
	\node[label=center:{$(c\,b\,d)\{d\}$}](cbd) at (0,6) {};
	\node[label=center:{$(c\,b\,b)\{b\}$}](cbb) at (2,5) {};
	\node[label=center:{$(c\,b\,a)\{a\}$}](cba) at (4,4) {};
	\node[label=center:{$(d\,b\,d)\{d\}$}](dbd) at (12,6) {};
	\node[label=center:{$(d\,b\,b)\{b\}$}](dbb) at (10,5) {};
	\node[label=center:{$(d\,b\,a)\{d\}$}](dba) at (8,4) {};
	
  %
  \tikzstyle{LabelStyle}=[fill=none]
	\tikzstyle{EdgeStyle}=[above,shorten >= 0.25cm, shorten <=0.25cm]
	  \path (ccd) edge [->,shorten >= 0.25cm, shorten <=0.25cm] (ccb);
	
	\path  (dcd) edge [->,shorten >= 0.25cm, shorten <=0.25cm] (dcb);
	\path  (cbd) edge [->,shorten >= 0.25cm, shorten <=0.25cm] (cbb);
	\path  (cbb) edge [ultra thick,->,shorten >= 0.25cm, shorten <=0.25cm] (cba);
	\path  (dbd) edge [->,shorten >= 0.25cm, shorten <=0.25cm] (dbb);

	\path  (dbb) edge [<-,shorten >= 0.25cm, shorten <=0.25cm] (dba);
	\path  (dcb) edge [ultra thick,<-,shorten >= 0.25cm, shorten <=0.25cm] (dca);

	\path  (ccb) edge [ultra thick,<-,shorten >= 0.5cm, shorten <=0.5cm] (dcb);
	\path  (cba) edge [ultra thick,->,shorten >= 0.5cm, shorten <=0.5cm] (dba);

	\path  (cbb) edge [ultra thick,<-,shorten >= 0.15cm, shorten <=0.15cm] (ccb);
	
	\path  (cba) edge [->,shorten >= 0.15cm, shorten <=0.15cm] (cca);
	\path  (dba) edge [ultra thick,->,shorten >= 0.15cm, shorten <=0.15cm] (dca);

	\path  (dca) edge [<-,bend left,shorten >= 0.2cm, shorten <=0.5cm] (dcd);
	\path  (cbd) edge [->,bend left,shorten >= 0.2cm, shorten <=0.5cm] (cba);
	
\end{tikzpicture}
}

%

%% file: result_tables.tex
\newcommand{\VV }{\cellcolor{green!25!white}}
\newcommand{\VX }{\cellcolor{rgb:green,0.1;gray,0.5;white,1}}

\newcommand{\XX }{\cellcolor{gray!65!white}}

\begin{table*}[t]
\small
\centering
\begin{tabular}{l||c|c|c|c|c}
Voting rule & FIP & FBRP & FDBRP & restricted-FIP & Weak-FIP \\
            
\hline
Dictator         & V \VV     & V \VV & - & V \VV & V\VV \\
\hline
Plurality (lex.) & X \XX     & X \XX (Ex.~\ref{ex:no_FBRP})  & V  \VV (Thm.~\ref{th:FDRP}) & V \VV & V \VV \\
Plurality (LD)   &X \XX (Ex.~\ref{ex:noFIP_r}) & ?     & V [M15] \VV &  V \VV  & V \VV \\
Plurality (SD)   &X \XX (Ex.~\ref{ex:noFIP_r}) & ?& V [RW12] \VV &  V \VV  & V \VV \\
Plurality (rand.) &X \XX (Ex.~\ref{ex:noFIP_r}) &  
 X \XX & X \XX & X \XX (Ex.~\ref{ex:no_RFIP_r}) &  V \VV   (Thm.~\ref{th:weakFDRP_r})\\
Weighted Plurality (lex.) &X \XX & X \XX & X \XX (Ex.~\ref{ex:no_RFDRP_w}) & ? & ? \\
\hline
Veto           &X \XX  & X \XX [M16] &  V \VV [RW12,LR12] &  V \VV &  V \VV  \\
$k$-approval ($k\geq 2$) &X \XX &  X  \XX [LR12,L15] &-& X \XX & X \XX  [M16]\\
Borda       &X \XX   & X \XX [RW12,LR12] &- & X\XX &  X \XX [RW12] \\
PSRs (except $k$-approval)&X \XX  &  X  \XX [LR12,L15] &-& ? & ? \\
Approval &X \XX & X \XX [M16] & - & V \VV [M16] & V \VV \\
Other common rules & X \XX  &  X  \XX [KLR16] &-& ? & ? \\
\end{tabular}
\caption{\label{tab:results_rules}
Positive results carry to the right side, negative to the left side. We assume lexicographic tie breaking in all rules except Plurality.
FDBRP is only well-defined for Plurality and Veto. Reference codes: RW12~\cite{reyhani2012best}, LR12~\cite{lev2012convergence}, M15~\cite{Meir15}, L15~\cite{Lev15}, M16~\cite{Meir:2016:COMSOC}, KLR16~\cite{koolyk2016convergence}. }
\end{table*}

%% file: plurality.TEAC.appendix.tex
\section{Proofs}

\begin{rlemma}{lemma:rel} Consider any cardinal utility function $u$ and the partial preference order $Q$ it induces on subsets by random tie-breaking.  
$Q$ holds Axioms~K+G+R. 
\end{rlemma}

\begin{proof}
Let $u$ be any utility scale, we will show that all axioms hold.
Let $a,b\in C$ and $W\subseteq C\setminus\{a,b\}$. 
$$u(\{a\} \cup W) = \frac{1}{|W|+1|}\(u(a) + \sum_{c\in W}u(c)\), u(\{b\} \cup W)=\frac{1}{|W|+1|}\(u(b) + \sum_{c\in W}u(c)\) = u(\{b\} \cup W),$$ thus $\{a\} \cup W \succ_Q \{b\} \cup W$, and Axiom~R holds.

Let $a\in C, W\subseteq C$ s.t. $\forall b \in W, a \ord b$. Then
\begin{align*}
u(a) &= \frac{1}{|W|+1}\(u(a)+\sum_{b\in W}u(a)\) > \frac{1}{|W|+1}\(u(a)+\sum_{b\in W}u(b)\)  = u(\{a\} \cup W)\\
& >\frac{1}{|W|+1}\(u(W)+\sum_{b\in W}u(a)\) = \frac{1}{|W|+1}u(W) + \frac{|W|}{|W|+1}u(W)= u(W),
\end{align*}
thus $a \succ_Q \{a\} \cup W \succ_Q W$ and Axiom~G holds. 

Axiom~K1 follows immediately from G. K2 also follows if preferences are strict. Even if there are ties, and $a\succeq w$ for all $a\in A,w\in W$ then:
$$u(A)\geq \min_{a\in A}u(a) \geq \max_{w\in W}u(w)\geq u(W),$$
i.e., $A \succeq_Q W$. 
\end{proof}

\begin{definition}\label{def:matching} 
Suppose that $X,Y\subseteq C$, $k=|X|\leq |Y|=K$. Sort $X,Y$ in increasing order by $Q$. Let $r_j = \ceil{\frac{j}{k}K}$. 
Partition $Y$ into sets $Y_1,\ldots,Y_k$ s.t. for $j<K$,  $Y_j=\{y_{r_{j-1}+1},\ldots, y_{r_j}\}$ (e.g., if $k=3,K=7$, then $Y$ is partitioned into $Y_1=\{y_1,y_2,y_3\}, Y_2=\{y_4,y_5\}, Y_3= \{y_6,y_7\}$).

$X$ \emph{match-dominates} $Y$ according to $Q$ if:
\begin{itemize}
	\item (I) $\forall j\leq k \forall y\in Y_j$, $x_j \succeq y$; and
	\item either (IIa) at least one relation is strict, or (IIb) $K \mod k\neq 0$.
\end{itemize}
If $|X|>|Y|$, then $X$ \emph{match-dominates} $Y$ if $Y$ match-dominates $X$ according to the reverse of $Q$. 
\end{definition} 
Intuitively, match-domination means that for any $q\in[0,1]$, there is a fraction $q$ of the set $X$ that  dominates a fraction of $1-q$ from the set $Y$: at least one $x\in X$ dominates all of $Y$, at least 20\% of $X$ dominate at least 80\% of $Y$, and so on. 

\begin{lemma}\label{lemma:SD}
Let $\vec a,\vec a'$ be two profiles that differ by a single vote, and define $X=f(\vec a),Y=f(\vec a')$.\footnote{Without some restriction on $X,Y$, the lemma is incorrect. E.g. if $x_1\succ y_1 \succ y_2 \succ x_2 \succ y_3 \succ y_4$, then $X$ stochastically dominates $Y$ but there is no way to derive $X\succ Y$ from the axioms K+G+R.}
 
The following conditions are equivalent for any strict order $Q$ over $C$:
\begin{enumerate}
	\item $X$ stochastically dominates $Y$ under preferences $Q$ and uniform lottery.
	\item The relation $X\succ Y$ is entailed by $Q$ and the Axioms~K+G+R and transitivity.
	\item  $u(X)>u(Y)$ for every $u$ that is consistent with $Q$.
	\item  $X$ match-dominates $Y$ according to $Q$. 
\end{enumerate} 	
\end{lemma}
\begin{proof}
The equivalence of (1) and (3) is immediate, and used e.g. in \cite{reyhani2012best}.

(2) $\Rightarrow$ (3). If $X\succ Y$ follows from the axioms, then there is a sequence of sets $X=X_0\succ X_1 \succ \cdots \succ X_k =Y$ such that each $X_j \succ X_{j+1}$ follows from a single axiom K,G, or R. Thus it is sufficient to show for $X\succ Y$ that follows from a single axiom. 

If $X\succ Y$ follows from Axiom~R, then $X=\{a\}\cup W,Y=\{b\}\cup W$ for some $W\subseteq C\setminus\{a,b\}$ and $a\succ b$. Thus
$$u(X) = u(\{a\} \cup W) = \frac{1}{|W|+1|}\(u(a) + \sum_{c\in W}u(c)\) > \frac{1}{|W|+1|}\(u(b) + \sum_{c\in W}u(c)\) = u(\{b\}\cup W)=u(Y).$$

If $X\succ Y$ follows from Axiom~G, then either $X=Y\cup \{a\}$ and $a\succ b$ for all $b\in Y$, or $X=\{x\}$ and $Y=\{x\}\cup W$ where $x\succ w$ for all $w\in W$. For the first case
\begin{align*}
u(X) &=\frac{1}{|Y|+1|}u(a) + \frac{1}{|Y|+1}\sum_{y\in Y}u(y) =\frac{1}{|Y|+1}\frac{1}{|Y|}\sum_{y\in Y}u(a) +   \frac{1}{|Y|+1}\sum_{y\in Y}u(y)\\
&> \frac{1}{|Y|+1|}\frac{1}{|Y|}\sum_{y\in Y}u(y) +   \frac{1}{|Y|+1|}\sum_{y\in Y}u(y)\\
&=\(1+\frac{1}{|Y|}\) \frac{1}{|Y|+1}\sum_{y\in Y}u(y) =  \frac{1}{|Y|}\sum_{y\in Y}u(y) = u(Y).
\end{align*}
For the second case, 
$$u(X) = u(x) = \frac{1}{|Y|}\sum_{y\in Y}u(x) = \frac{1}{|Y|}\(u(x) +\sum_{w\in W}u(x)\) >  \frac{1}{|Y|}\(u(x) +\sum_{w\in W}u(w)\) = u(Y).$$
 
If $X\succ Y$ follows from Axiom~K, then $u(x)>u(y)$ for any $x\in X,y\in Y$ which is a trivial case.


(3) $\Rightarrow$ (4). 
Suppose that $u(X)>u(Y)$ for all $u$. Suppose first $|X|\leq |Y|$. If $|X|$ does not match-dominate $Y$ then either (I) there  is an element $x_{j'}$ that is less preferred than some element $y'\in Y_{j'}$; or (II) for all $j$ and all $y\in Y_j$, $x_j=_Q y$ and $|Y_j|=\frac{K}{k}=q$ for all $j$. We will derive a contradiction to (3) in either case. In the latter case, we have $u(x_j)=u(Y_j)$ for all $j$ and thus 
$$u(Y)=\frac{1}{K}\(\sum_{j\leq k}|Y_j|u(Y_j) \)=\frac{\sum_{j\leq k}q u(x_j) }{K} = \frac{\sum_{j\leq k}q u(x_j)}{kq} =  u(X),$$
In contradiction to (3).

Thus we are left with case (I). That is, there are $j' \leq k$ and $y'\in Y_{j'}$ s.t. $x_{j'} \prec y'$.  We define the (possibly empty) set $X'\subseteq X$ as all elements $\{x: x\succ x_{j'}\}$. We define $Y'\subseteq Y$ as $\{y: y\succeq y'\}$. By construction, for any $j>j'$, $Y_j\subseteq Y'$. Thus
 $$|Y'|\geq 1+\sum_{j=j'+1}^k|Y_j|=1+\sum_{j=j'+1}^k(r_j-r_{j-1})=(K-r_{j'})+1 = (K-\ceil{\frac{j'}{k}K})+1> K-\frac{j'}{k}K=K(1-\frac{j'}{k}),$$
whereas $|X'|\leq k-j'$. 
We define $u$ as follows: $u(x)=1,u(y)=1$ for all $x\in X',y\in Y$, and $u(z)=0$ for all other elements. Note that $X',Y'$ contain the top elements of $X,Y$, respectively. In addition, $y'$ is the minimal element in $Y'$ and by transitivity $y'\succ x$ for all $x\in X\setminus X'$. Thus $u$ is consistent with $Q$.

We argue that $u(Y)>u(X)$ in contradiction to (3). Indeed, $u(X) = \frac{|X'|}{|X|}\leq \frac{k-j'}{k}=1-\frac{j'}{k}$. 
$$u(Y) = \frac{|Y'|}{|Y|} > \frac{(1-\frac{j'}{k})K}{K} = 1-\frac{j'}{k} = \frac{k-j'}{k} \geq \frac{|X'|}{|X|}= u(X),$$
so we get a contradiction to (3) again.  
Thus $X$ matching-dominate $Y$.

(4) $\Rightarrow$ (2).
This is the only part of the proof where we use the profiles from which $X,Y$ are obtained.
When a single voter moves, either the winner set changes by a single candidate (added, removed, or swapped), or $X$ is a single candidate, or $Y$ is a single candidate. We prove case by case.
\begin{itemize}
\item The case where $|X|=|Y|=1$ is immediate. 
	\item Suppose $|X|=1$ (i.e. $X=\{x\}$) and $|Y|=K>1$. Then $X$ match-dominates $Y$ means that $x \succeq y$ for all $y\in Y$, with at least one relation being strict, w.l.o.g. $y_K$ (least preferred in $Y$). Then $X \succeq \{y_1,\ldots,y_{K-1}\} \succ Y$, where the first transition is by Axiom~K2 and the second is by Axiom~G.
	\item The case of $|Y|=1$ is symmetric.
	\item Suppose $|X|=|Y|=k$. Then  $X$ match-dominates $Y$ means that $x_i\succeq y_i$ for all $i$. For all $t\in\{0,1,\ldots,k\}$, let $X^t=\{x_1,\ldots,x_t,y_{t+1},\ldots,y_k\}$. Then $X^{t-1} = X^{t}$ if $x_t=y_t$, and $X^{t-1} \succ X^{t}$ otherwise from Axiom~R. In addition, $X=X^0,Y=X^k$ thus $X\succ Y$ from transitivity. 
	\item Suppose $|X|=k,|Y|=k+1$. Then $X$ match-dominates $Y$ means that $|Y_1|=\ceil{\frac{k+1}{k}}=2$, and all other sets $Y_j$ are singletons $Y_j=y_j$. Consider the set $Y'$ that includes the top $k$ elements of $Y$. Since $x_1$ is (weakly) preferred to both candidates in $Y_1$, $Y'$ is match-dominated by $X$. By the previous bullet $X \succeq Y'$ follows from Axiom~R and transitivity. Finally, $Y'\succ Y=Y'\cup\{\min Y\}$ by Axiom~G.
\end{itemize}
\end{proof}

The following is an immediate corollary:

\begin{rproposition}{th:SD_axioms}
 A step $\vec a\step i \vec a'$ is a better-response under random tie-breaking and stochastic dominance, if and only if $f(\vec a') \succ_i f(\vec a)$ is entailed by $Q_i$, the Axioms~K+G+R, and transitivity.
\end{rproposition}

\begin{rproposition}{th:LD_axioms}
 A step $\vec a\step i \vec a'$ is a better-response under unknown tie-breaking and local dominance, if and only if $f(\vec a') \succ_i f(\vec a)$ is entailed by $Q_i$,  Axioms~K+G, and transitivity.
\end{rproposition}
\begin{proof}
Suppose that $X=f(\vec a')$ locally-dominates $Y=f(\vec a)$. Let $Z=X\cap Y$, and $X'=X\setminus Z$, $Y'=Y\setminus Z$. We must have $x \succ_i y$ for any $x\in X, y\in Y'$, otherwise, a tie-breaking order that selects $y$ first and $x$ second would make $i$ strictly lose when moving from $Y$ to $X$. Similarly, $x \succ_i y$ for any $x\in X',y\in Y$. If $Z=\emptyset$ then $X=X' \succ_i Y'= Y$ follows from Axiom~K. Otherwise,  by repeatedly applying Axiom~G we get $X \succeq_i Z \succeq_i Y$ with at least one relation being strict. 

In the other direction,  since Axiom~G can only be used to add elements lower (or higher) than all existing elements, it may only induce relations of the form $Z\succ Z \cup Y'$ where $z \succ y$ for all $z\in Z,y\in Y'$; or relations of the form $Z \cup X' \succ Z $ where $x \succ z$ for all $z\in Z,x\in X'$. Thus if $X\succ Y$ follows from Axiom~G, they must have the form $X=Z\cup X', Y=Z\cup Y'$ where $x\succ z \succ y$ for all $x\in X',z\in Z,y\in Y'$. To see that this entails local dominance, let $x_L=L(X)$ be the first element in $X$ according to order $L\in \pi(C)$, and likewise for $Y$.  For any $L$, $x_L \succeq y_L$ (with equality iff $L(X)=L(Y)\in Z$). Further, either $X'$ or $Y'$ are non-empty (w.l.o.g. $X'$). Consider an order $L'$ such that $L'(X)\in X'$, then $x_{L'} \succ y$ for all $y\in Y$ and in particular $x_{L'}\succ y_{L'}$. 
\end{proof}